%% file: paper.tex
\def\preprint{1}

\ifdefined\preprint
  \documentclass[sigconf,nonacm]{acmart}
\else
  \documentclass[sigconf,review,anonymous]{acmart}
\fi

\usepackage{microtype}
\usepackage{graphicx}
\usepackage{subcaption}
\usepackage{booktabs}
\usepackage[table]{xcolor}
\usepackage{framed}

\definecolor{lightgrey}{gray}{0.95}

\usepackage{xurl}

\usepackage{amsmath}
\usepackage{mathtools}
\usepackage{amsthm}

\usepackage{listings}

\ifdefined\preprint
\else
\usepackage{minted}
\fi

\usepackage{xcolor}
\usepackage{fancyvrb}
\usepackage{verbatimbox}

\definecolor{lstkeyword}{RGB}{0,0,120}
\definecolor{lstcomment}{RGB}{96,96,96}
\definecolor{lststring}{RGB}{160,32,32}
\definecolor{lstbackground}{RGB}{248,248,248}

\lstdefinestyle{icmlpython}{
  language=Python,
  backgroundcolor=\color{lstbackground},
  basicstyle=\ttfamily\scriptsize,
  keywordstyle=\color{lstkeyword}\bfseries,
  commentstyle=\color{lstcomment}\itshape,
  stringstyle=\color{lststring},
  numbers=left,
  numberstyle=\tiny\color{lstcomment},
  stepnumber=1,
  numbersep=6pt,
  showstringspaces=false,
  tabsize=2,
  breaklines=true,
  breakatwhitespace=true,
  frame=single,
  rulecolor=\color{black},
  framerule=0.4pt,
  framesep=4pt,
  captionpos=b,
  escapeinside={/*@}{@*/},
}

\newcommand{\code}[1]{\lstinline[basicstyle=\ttfamily]!#1!}

\usepackage[capitalize,noabbrev]{cleveref}

\theoremstyle{plain}

\theoremstyle{definition}

\theoremstyle{remark}

\newcommand{\inspect}{Inspect}

\newcommand{\intercode}{Intercode}

\newcommand{\cybench}{Cybench}
\newcommand{\tool}{Evolve-CTF}
\newcommand{\intercodepythonchallengesavailable}{20}
\newcommand{\cybenchpythonchallengesavailable}{18}
\newcommand{\intercodepythonchallengessolvable}{19}
\newcommand{\cybenchpythonchallengessolvable}{2}

\newcommand{\totalcybenchchallenges}{2}
\newcommand{\totalchallenges}{16}
\newcommand{\instancesperfamily}{24}
\newcommand{\tokenbudget}{200,000}
\newcommand{\repeats}{5}

\newcommand{\totalmodels}{9}
\newcommand{\totalmodelconfigurations}{13}

\newcommand{\ShortClaude}{Claude}
\newcommand{\ShortDeepSeek}{DS}
\newcommand{\ShortGPT}{GPT}
\newcommand{\ShortGPTReasoning}{GPT-R}
\newcommand{\ShortGemini}{Gem}
\newcommand{\ShortGeminiReasoning}{Gem-R}
\newcommand{\ShortGrok}{Grok}
\newcommand{\ShortKimi}{Kimi}
\newcommand{\ShortKimiReasoning}{Kimi-R}
\newcommand{\ShortMiniMax}{Mini}
\newcommand{\ShortMagistral}{Mag}
\newcommand{\ShortQwen}{Qwen}
\newcommand{\ShortQwenReasoning}{Qwen-R}

\newcommand{\claudeopus}{Anthropic Claude Opus 4.5}
\newcommand{\deepseekthree}{DeepSeek V3.2}
\newcommand{\geminithree}{Google Gemini 3 Pro}
\newcommand{\gptfive}{OpenAI GPT-5.1}
\newcommand{\grokfour}{xAI Grok 4}
\newcommand{\kimiktwo}{Moonshot AI Kimi K2}
\newcommand{\mistralmagistral}{Mistral Magistral Med.\ 3}
\newcommand{\minimax}{MiniMax M2.1}
\newcommand{\qwenthree}{Qwen 3 Max}

\definecolor{lightgray1}{gray}{0.95}
\definecolor{lightgray2}{gray}{0.90}

\newcommand{\mypara}[1]{\smallskip\noindent\textbf{#1}\ }

\newcommand{\myparaflowing}[1]{\smallskip\noindent\textbf{#1}}

\usepackage[textsize=tiny]{todonotes}

\acmConference[CONF '26]{Some conference}{2026}{To be announced}

\title{Capture the Flags: Family-Based Evaluation of Agentic LLMs via Semantics-Preserving Transformations}

\author{Shahin Honarvar}
\affiliation{%
  \institution{Department of Computing, Imperial College London}
  \country{UK}
}
\email{anon@anon.anon}

\author{Amber Gorzynski}
\affiliation{%
  \institution{Department of Computing, Imperial College London}
  \country{UK}
}

\author{James Lee-Jones}
\affiliation{%
  \institution{Department of Computing, Imperial College London}
  \country{UK}
}

\author{Harry Coppock}
\affiliation{%
  \institution{AI Security Institute}
  \country{UK}
}
\affiliation{%
  \institution{Department of Computing, Imperial College London}
  \country{UK}
}

\author{Marek Rei}
\affiliation{%
  \institution{Department of Computing, Imperial College London}
  \country{UK}
}
\email{anon@anon.anon}

\author{Joseph Ryan}
\affiliation{%
  \institution{Royal Grammar School Guildford}
  \country{UK}
}

\author{Alastair F. Donaldson}
\affiliation{%
  \institution{Department of Computing, Imperial College London}
  \country{UK}
}

\begin{document}

\begin{abstract}
Agentic large language models (LLMs) are increasingly evaluated on cybersecurity tasks using capture-the-flag (CTF) benchmarks, yet existing pointwise benchmarks offer limited insight into agent robustness and generalisation across alternative versions of the source code.
We introduce \emph{CTF challenge families}, whereby a single CTF is used to generate a family of semantically-equivalent challenges via semantics-preserving program transformations, enabling controlled evaluation of robustness while keeping the underlying exploit strategy fixed.
We present \tool{}, a tool that generates CTF families from Python challenges using a range of transformations.
Using \tool{} to derive families from \cybench{} and \intercode{} challenges, we evaluate \totalmodelconfigurations{} agentic LLM configurations with tool access.
We find that models are remarkably robust to renaming and code insertion, but that composed transformations and deeper obfuscation degrade performance by requiring more sophisticated tool use.
Enabling explicit reasoning has little effect on success rates.
Our work contributes a technique and tool for future LLM evaluations, and a large dataset characterising the capabilities of current state-of-the-art models in this domain.
\end{abstract}

\maketitle

\section{Introduction}

Agentic LLMs are increasingly deployed in cybersecurity settings, ranging from vulnerability discovery and exploit generation to automated penetration testing and defensive analysis~\citep{Ferrag2025, Zhang2025}. As these systems gain autonomy and access to external tools, evaluating their tool-based cyber-capabilities (rather than isolated coding or reasoning skills) is increasingly important.
Recent work has converged on \emph{capture-the-flag challenges} (CTFs) as a natural testbed for this purpose~\citep{Chapman2014,Valdemar2021,Yang2023,Zou2024,Debenedetti2024,AndyZhang2025,Ji2025,Bakker2025}. A CTF is a program, binary, or system containing a deliberately-introduced vulnerability that can be exploited to recover a secret value, known as a \emph{flag}.
Solving a CTF typically requires a combination of code comprehension, modification and tool use.

Existing CTF-based benchmarks for LLMs, such as \cybench{}~\citep{AndyZhang2025} and \intercode{}~\citep{JohnYang2023}, provide a valuable starting point by standardising evaluation on curated collections of CTFs. However, the challenges in these benchmarks are all independent tasks.
As a result, evaluation outcomes are \emph{pointwise}:
they reveal how a model performs on a fixed set of heterogeneous tasks,
but offer limited insight into robustness against input perturbations,
or the ability of models to \emph{generalise}.
Since CTF benchmarks are often public, or drawn from public competitions, their use in evaluations may suffer from dataset contamination, similar to effects observed when applying LLMs to mathematical problems~\cite{Zhang2024}.

We propose a complementary approach to evaluation based on \emph{CTF challenge families}.
Given an original CTF we apply semantics-preserving but obfuscating program transformations to generate a family of related challenges.
All challenges in a family (henceforth called \emph{instances}) share the vulnerability associated with the original challenge, thus all are ultimately solvable via the same underlying exploit strategy.
However, instances vary widely in syntactic structure, control flow, meaningfulness of variable/function/class names and degree of obfuscation.
Unlike fixed benchmarks, CTF families enable systematic evaluation of agentic LLM performance across structured spaces of related challenges, helping disentangle genuine reasoning ability from pattern matching.
CTF families make it easier to investigate whether an LLM really understands a given challenge and can solve it even after superficial modifications, rather than relying on confounding cues and memorisation.

We emphasise that CTF families do not aim to test new cybersecurity competencies.
Rather, they test whether an agent that can solve a CTF truly possesses the underlying code reasoning and tool-use capability, or whether its success depends on surface-level cues such as recognisable identifier names or familiar code structure.

We put the idea of CTF families into practice through a new tool, \tool{}.
Given a Python-based CTF, \tool{} automatically generates a CTF family by applying a range of transformations (detailed in \Cref{sec:transformationdesign}).
We present a large-scale evaluation over CTF families generated from \totalchallenges{} Python challenges from \cybench{} and \intercode{}, evaluating \totalmodelconfigurations{} LLM configurations spanning \totalmodels{} base models, with agentic capabilities enabled through the \inspect{} framework~\cite{Inspect2024}. Our key findings are:

\mypara{LLMs are highly robust to identifier renaming and isolated source code transformations}
Across all evaluated models, performance is largely unchanged by identifier renaming or injecting a single class of redundant code (e.g., extraneous loops or unused functions).
In contrast, robustness degrades sharply when transformations are \emph{composed}.

\mypara{Composite transformations lead to increased tool use}
When challenges involve composed transformations, models make substantially more tool calls.
Transcript analysis reveals strategic tool use such as searching for security-relevant keywords and iteratively inspecting surrounding code regions.

\mypara{LLMs are hindered by deeper, tool-based obfuscation}
Models rarely solve challenges obfuscated with PyObfuscator, a third-party code obfuscator.
Transcript analysis shows that models sometimes attempt, with limited success, to write Python scripts that partially reverse the obfuscation, occasionally enabling them to then solve the challenge.

\mypara{Explicit reasoning has little impact}
For models that support toggling explicit reasoning, we find that setting reasoning to high vs.\ using default minimal reasoning settings has little impact on the model's ability to solve CTF families.

\mypara{Some benchmark CTFs lack discriminative power}
Three CTFs from the \intercode{} suite are so simple that nearly all evaluated models solve \emph{all} instances of CTF families generated from them.
These challenges thus provide little signal for distinguishing state-of-the-art agentic LLMs.
This highlights an additional use case for \tool{}: assessing the difficulty and discriminative value of existing benchmarks.

Overall, CTF families provide a new way to evaluate the performance of agentic LLMs on cybersecurity tasks that goes beyond single-instance benchmarks, offering a principled way to probe robustness, generalisation and tool usage.

\ifdefined\preprint
\else
\mypara{Tool and data availability}
Following the review process we plan to make \tool{} available as open source under the Apache 2 license, together with all experimental data.
\fi

\section{Background}

\begin{figure}
\includegraphics{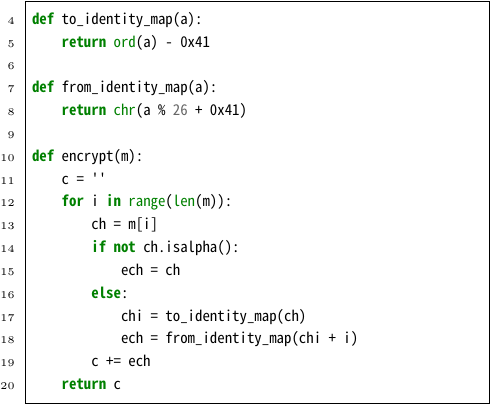}
\caption{Excerpt from the \emph{Dynastic} CTF from \cybench{}}\label{fig:dynastic}
\end{figure}

\mypara{CTF challenges}\label{sec:ctfs} Capture-the-flag challenges (CTFs) are a class of security-oriented tasks in which participants must analyze a system or artifact to recover a hidden secret,
known as the flag.
\Cref{fig:dynastic} presents a simple example, \emph{Dynastic}, drawn from \cybench{}~\cite{AndyZhang2025}.
In this challenge, the participant is given the code used to encrypt a message (\Cref{fig:dynastic}) and an encrypted message containing the flag (omitted for brevity).
The goal is to determine how to invert the encryption procedure, thereby recovering the original message and revealing the flag.
Two suites of CTFs that are widely-used for LLM-based evaluations, and which we use in this work, are \intercode{}~\citep{JohnYang2023} and \cybench{}~\citep{AndyZhang2025}.

\mypara{The \inspect{} framework}
\inspect{}~\citep{Inspect2024} is an open-source framework for standardised evaluation of agentic LLMs, supporting configurable tasks and models and recording detailed evaluation logs.
We base our evaluation on \inspect{} as it already incorporates the \cybench{} and \intercode{} suites and integrates with a wide range of model APIs.
\ifdefined\preprint
\else
File \texttt{DS\_Orig.txt} in our artifact~\cite{ARTIFACT} shows the transcript for an LLM (DeepSeek) solving a \cybench{} CTF via \inspect{}.
\fi

\section{Transformations Supported by \tool{}}\label{sec:transformationdesign}

A CTF family comprises many \emph{instances}: the original CTF plus follow-on challenges derived via transformations.
Our tool, \tool{}, generates such families from Python challenges,
confirming (via each challenge's golden solution) that generated instances remain solvable.
We focus on Python because it is widely used~\citep{top-programming-languages-2025},
LLMs perform well on it~\citep{llms-love-python},
and many existing CTFs use Python.
Conceptually, our approach is language-agnostic and could support other languages with sufficient engineering.
We integrate \tool{} with \inspect{} so a wide range of models can be applied to generated instances.

We now describe the transformations supported by \tool{}.
\ifdefined\preprint
\else
The files \texttt{dynastic\_*.py} in our artifact illustrate these by showing the effect of each \tool{} transformation on the \emph{Dynastic} CTF.
\fi
All transformations are \emph{semantics-preserving}, so the exploit strategy for the original challenge applies to every instance in a family.
Source code manipulation performed in $R$ and $T_1$---$T_5$ is performed using LibCST, a concrete syntax tree library for Python that preserves formatting and comments~\cite{libcst}.

The transformations probe LLM robustness across a spectrum,
from coarse obfuscation to more sophisticated semantics-preserving rewrites inspired by compiler-testing techniques~\cite{DBLP:journals/pacmpl/DonaldsonELT17,DBLP:conf/pldi/LeAS14}.
The modular design of \tool{} also facilitates integrating additional transformations in future work.

\myparaflowing{$R$: Rename identifiers} replaces the names of all variables, functions and classes with randomly-chosen alternatives
drawn from two sources: coding-related vocabulary (concatenated programming terms from a curated word list) and multilingual random strings generated across various natural languages including mixed-language combinations.
In the case where replacement identifiers are instead randomly-generated, randomisation is used to determine both the length of replacement identifiers (sampled from a configurable range) and their composition.

\myparaflowing{$T_1$: Insert loops} introduces redundant nested loops at randomly-sampled eligible program locations. At each location, a nesting depth up to a configurable maximum is chosen, with each level selecting a \code{for} or \code{while} loop uniformly at random. By construction, the conditions and iteration ranges of outermost loops are guaranteed to evaluate to false and be empty, respectively, ensuring no semantic impact. However, the expressions used for conditions and ranges can be complex, so that these guarantees are far from evident without careful analysis. Variables for \code{for} loops are generated using either multilingual random strings, with lengths sampled from a configurable range, or are drawn from coding-related vocabulary (similar to fresh identifiers used by $R$). In the case of inner loops, variable names are also sampled, with a configurable probability, from names occurring in the original challenge code. Reusing original program names inside loop nests has the potential to make the transformed code more confusing from the model's point of view, without risking semantic change since such inner loops are dynamically unreachable. Loop body statements are generated in a randomised fashion, and may include e.g.\ bitwise operations, arithmetic, string manipulations, and nested comprehensions.

\myparaflowing{$T_2$: Insert conditionals} introduces redundant control-flow constructs at randomly sampled eligible locations (as in $T_1$).
At each location, either a nested \code{if} statement with a provably false outer condition and random assignments is inserted,
or the existing code is wrapped in a (possibly nested) \code{try}-\code{except} block.
Similar to the case of loop insertion the guards of outer-most conditionals are guaranteed to evaluate to false, while nested conditionals (being unreachable) are unrestricted and can make use of identifiers from the original challenge as well as randomly-generated multilingual identifiers.
As with loops, nested conditionals may refer to variable names from the original challenge.

\myparaflowing{$T_3$: Insert functions} introduces redundant callable definitions at randomly sampled eligible locations (as in $T_1$).
Each location probabilistically receives either a nested \code{def} or lambda assignment, with nesting depth and parameter lists sampled at random up to configurable maxima.
Newly-introduced functions can have randomly-generated names or names deliberately chosen to be similar to names in the enclosing context.
Names are chosen to avoid clashes, using fresh identifiers at outer levels while safely reusing in-scope names (including those from the original challenge) at inner levels.
Function and lambda parameter names may also be drawn from names used elsewhere in the challenge.

\myparaflowing{$T_4$: Insert code comments} adds synthetic comments at randomly sampled eligible locations (as in $T_1$), preserving indentation.
Comment content is chosen probabilistically: with configurable probability from a curated pool of natural English programming statements, otherwise as meaningless multilingual strings of random lengths.

\myparaflowing{$T_5$: Combine $T_1$--$T_4$} applies the four sub-transformations sequentially---loops ($T_1$), conditionals ($T_2$), functions ($T_3$), and comments ($T_4$)---allowing later transformations to operate on code introduced earlier.
Each transformation receives an equal fixed insertion budget based on the original eligible locations, limiting code growth and keeping overall size comparable to applying a single $T_i$ in isolation.
Comments are inserted last, while the ordering of the other transformations is arbitrary but sensible, enabling, for example, functions within added loops or conditionals without incurring a combinatorial explosion of orderings.

\myparaflowing{$O$: Apply PyObfuscator} applies more aggressive obfuscation using the PyObfuscator tool~\cite{pyobfuscator}.
We use a medium obfuscation level that renames all identifiers, removes docstrings, encrypts string literals, and compresses the resulting code with gzip.

Although both $R$ and $O$ rename identifiers, their purposes are distinct:
$R$ performs only renaming, enabling isolated measurement of sensitivity to identifiers,
while $O$ applies renaming as part of a much more aggressive transformation pipeline that also encrypts strings and compresses code.

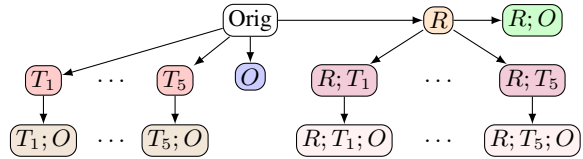
\begin{figure}[t]
\centering
\begin{tikzpicture}[
    >=latex,
    every node/.style={draw, rounded corners, inner sep=2pt, font=\small},
    nodeBlue/.style={fill=blue!20},
    nodeOrange/.style={fill=orange!20},
    nodeGreen/.style={fill=green!20},
    nodeRed/.style={fill=red!20},
    nodePurple/.style={fill=purple!20},
    nodeBrown/.style={fill=brown!20},
    nodePink/.style={fill=pink!20},
    nodeTeal/.style={fill=teal!20},
    ellipsis/.style={draw=none}
]

\node (orig) at (1.75,0) {Orig};

\node[nodeBlue] (o) at (1.75,-0.75) {$O$};
\node[nodeOrange] (r) at (4.25,0) {$R$};
\node[nodeGreen] (ro) at (5.5,0) {$R;O$};

\draw[->] (orig) -- (r);
\draw[->] (orig) -- (o);
\draw[->] (r) -- (ro);

\node[nodeRed] (T1)   at (-1,-0.8) {$T_1$};
\node[ellipsis] (dots) at (-0.1,-0.8) {\(\mathinner{\ldots}\)};
\node[nodeRed] (T5)   at (0.75,-0.8) {$T_5$};

\draw[->] (orig) -- (T1);
\draw[->] (orig) -- (T5);

\node[nodePurple] (RT1)   at (3,-0.8) {$R;T_1$};
\node[ellipsis] (dots) at (4.25,-0.8) {\(\mathinner{\ldots}\)};
\node[nodePurple] (RT5)   at (5.5,-0.8) {$R;T_5$};

\draw[->] (r) -- (RT1);
\draw[->] (r) -- (RT5);

\node[nodeBrown] (T1O)   at (-1,-1.6) {$T_1;O$};
\node[ellipsis] (dots) at (-0.1,-1.6) {\(\mathinner{\ldots}\)};
\node[nodeBrown] (T5O)   at (0.75,-1.6) {$T_5;O$};

\draw[->] (T1) -- (T1O);
\draw[->] (T5) -- (T5O);

\node[nodePink] (RT1O)   at (3,-1.6) {$R;T_1;O$};
\node[ellipsis] (dots) at (4.25,-1.6) {\(\mathinner{\ldots}\)};
\node[nodePink] (RT5O)   at (5.5,-1.6) {$R;T_5;O$};

\draw[->] (RT1) -- (RT1O);
\draw[->] (RT5) -- (RT5O);

\end{tikzpicture}
\caption{Transforming an original CTF (Orig) leads to a CTF family, based on our five kinds of transformation ($T_1$--$T_5$), optionally preceded by renaming ($R$) and/or followed by obfuscation ($O$)}\label{fig:ctffamily}
\end{figure}

\mypara{CTF families}
From a given CTF, we generate a family of variants by applying any of $T_1$--$T_5$, optionally preceded by identifier renaming ($R$) and optionally followed by obfuscation ($O$); $R$ and $O$ may also be applied in isolation or directly composed.
Renaming applies only to identifiers in the original challenge, as renaming those introduced by transformations is uninformative, while $O$ is applied last since PyObfuscator compresses code and precludes further transformations.
Overall, an \tool{} family comprises \instancesperfamily{} instances (including the original), forming a tree (\Cref{fig:ctffamily}) in which each node is derived by applying a transformation to an existing instance (e.g., `$R;T_5;O$' from `$R;T_5$').
Since $T_5$ combines the effects of $T_1$--$T_4$, compositions such as $T_3;T_1;T_2$ do not appear in \Cref{fig:ctffamily}.
The tree structure serves two purposes: it controls combinatorial explosion (all possible combinations of $R$, $T_1$--$T_5$ and $O$ would yield hundreds of variants per CTF),
and it enables \emph{compositional analysis}---by comparing a node with its parent, we can isolate the marginal effect of each transformation applied on top of prior transformations.

\section{Experimental Design}

In \Cref{sec:results} we report a large-scale evaluation using CTF families generated by \tool{}, assessing \totalmodelconfigurations{} LLM configurations via the \inspect{} agentic framework. Here, we describe our selection of CTFs (\Cref{sec:choiceofctfs}), models (\Cref{sec:llms-under-eval}) and evaluation protocol (\Cref{sec:evaluation-protocol}).

\subsection{CTFs Used for Family Generation}\label{sec:choiceofctfs}
Focusing on the widely used \cybench{} and \intercode{} suites (already integrated into \inspect{}), we selected the CTFs summarised in \cref{tab:challenges} for evaluation (using the process detailed below). We group these CTFs into three categories: \emph{Decrypt} (writing and executing non-trivial decryption code), \emph{Modify} (modifying and executing provided code artifacts), and \emph{Exec} (executing provided scripts and interpreting results). This coarse-grained categorisation is intended only to aid visualisation of results in \Cref{sec:results}.

We selected these from the available \cybench{} and \intercode{} Python CTFs as follows.

We studied the \cybench{} and \intercode{} CTFs, determining that they contained \cybenchpythonchallengesavailable{} and \intercodepythonchallengesavailable{} challenges with a Python-based solution, respectively.
Because \tool{} \emph{increases} the source code complexity of challenges, to meaningfully assess LLM robustness it was important to select challenges that state-of-the-art models \emph{could} already solve (we would learn little from applying our transformations to already intractable challenges).

To make this selection, we applied three state-of-the-art models at time of CTF selection---Claude Sonnet 3.7, Claude Sonnet 4, o4 mini and o3, and found that \cybenchpythonchallengessolvable{} of the \cybench{} challenges and \intercodepythonchallengessolvable{} of the \intercode{} challenges could be reliably solved by these models within three repeat runs with a per-run token limit of \tokenbudget{}.
The division of challenges was stark: the solvable challenges could solved by all models well within this token limit, while the remaining Python-based challenges were almost never solved within these resource constraints by any of these models.

We had to further remove a number of \intercode{} challenges,
either because their files contained intentional syntax errors that prevented parsing and mutation by \tool{},
or due to problems with their provided golden solutions.
These problems
(in one case a solution bug, which we reported to the \intercode{} maintainers, in the other case the lack of programmatic solutions)
made it impossible to automatically confirm whether the transformations applied by \tool{} preserved the underlying exploit.

We note that recent work has shown high solve rates on \intercode{} with plain agents~\citep{Turtayev2024}, consistent with our own finding (\Cref{sec:results}) that several \intercode{} challenges lack discriminative power.
We retain these challenges precisely because our family-based evaluation can diagnose \emph{which} challenges and transformations remain informative.

\begin{table}[t]
  \caption{The Python CTFs used in our evaluation.
The \cybench{} (`C') \emph{Dynastic} and \emph{Primary Knowledge} challenges are denoted `D' and `PK', respectively, while \intercode{} (`I') challenges are numbered as per the \intercode{} benchmark.}
  \label{tab:challenges}
  \centering
  \small
  {
  \setlength{\tabcolsep}{0.75\tabcolsep}
  \begin{tabular}{@{} l l p{0.57\columnwidth} @{}}
    \textbf{Challenge} & \textbf{Category} & \textbf{Solution Description} \\
    \rowcolor{lightgray1}
    C-D              & Decrypt           & Reverse basic encryption algorithm \\
    \rowcolor{lightgray1}
    C-PK             & Decrypt           & Perform RSA decryption given RSA parameters $n$, $e$, and $c$ \\
    \rowcolor{lightgray1}
    I-79            & Decrypt            & Perform RSA decryption thrice given $e$, $c$ and modulus keys $n_1$, $n_2$, $n_3$ \\
    \rowcolor{lightgray2}
    I-00             & Modify            & Replace a call to \texttt{exec} with a call to \texttt{print} \\
    \rowcolor{lightgray2}
    I-13             & Modify            & Derive a missing portion of the flag from a static check in the source file \\
    \rowcolor{lightgray2}
    I-25             & Modify            & Execute a script which yields a decimal; provide its binary representation to obtain the flag \\
    \rowcolor{lightgray2}
    I-[30--34]        & Modify           & Derive a password from hardcoded check or remove the check \\
    \rowcolor{lightgray2}
    I-36             & Modify            & Modify the provided file to call the function \texttt{print\_flag} \\
    \rowcolor{lightgray2}
    I-78             & Modify            & Comment out a \texttt{sys.exit} call \\
    \rowcolor{lightgray1}
    I-05             & Exec              & Use a given script to decrypt flag \\
    \rowcolor{lightgray1}
    I-06             & Exec              & Apply a script to a given flag text file and enter a password when prompted \\
    \rowcolor{lightgray1}
    I-24             & Exec              & Execute a provided Python file \\
  \end{tabular}
  }
\end{table}

\subsection{LLMs Under Evaluation}\label{sec:llms-under-eval}
\Cref{tab:models} summarises the LLMs we evaluate: recent models spanning multiple providers and parameter scales (where disclosed).
We henceforth refer to each model by its short name.

Motivated by prior work showing that increased reasoning can degrade performance on tasks with distracting information~\citep{Gema2025}, we additionally evaluate high-reasoning configurations of \ShortGemini{}, \ShortGPT{}, \ShortKimi{}, and \ShortQwen{}.
These configurations are denoted by the `-R' suffix in short names (e.g., \ShortGPT{} vs.\ \ShortGPTReasoning{}).
We collected results for \ShortDeepSeek{} with high reasoning, but a tool-calling bug~\citep{deepseek-bug} rendered them invalid; we thus exclude this configuration.
\ShortClaude{} supports reasoning via a configurable token budget rather than discrete modes.
Due to the absence of a provider-endorsed notion of high reasoning, as well as financial constraints, we do not evaluate reasoning variants of \ShortClaude{}.
The remaining models (\ShortMagistral{}, \ShortGrok{}, \ShortMiniMax{}) do not expose reasoning controls and are evaluated in their default configurations.
This leads to \totalmodelconfigurations{} model configurations overall.
All models are evaluated at their default temperature settings.
Although analysing robustness across CTF families under varying temperatures would be informative, it would substantially increase the scale of an already large evaluation (\totalchallenges{} CTF families $\times$ \instancesperfamily{} instances $\times$ \totalmodelconfigurations{} model configurations), exceeding our available resources.

\begin{table}[t]
  \caption{The LLMs we evaluate, ordered by release date}
  \label{tab:models}
  \centering
  \small
  \begin{tabular}{@{} l l l l @{}}
    \toprule
    \textbf{Short} & \textbf{Model}                        & \textbf{Params} & \textbf{Released} \\
    \midrule
    \ShortDeepSeek{}    & \deepseekthree{}     &  685B           & 2025/12 \\
    \ShortMiniMax{}     & \minimax{}           &   229B          & 2025/12 \\
    \ShortClaude{}      & \claudeopus{}              &   Not public    & 2025/11  \\
    \ShortGemini{}      & \geminithree{}       &   Not public    & 2025/11 \\
    \ShortGPT{}         & \gptfive{}           &   Not public    & 2025/11 \\
    \ShortQwen{}        & \qwenthree{}         &   Not public    & 2025/09 \\
    \ShortKimi{}        & \kimiktwo{}          &   1T            & 2025/07 \\
    \ShortGrok{}        & \grokfour{}          &   Not public    & 2025/07 \\
    \ShortMagistral{}   & \mistralmagistral{}  &   Not public    & 2025/06 \\
    \bottomrule
  \end{tabular}
\end{table}

\subsection{Evaluation Protocol}\label{sec:evaluation-protocol}

To account for the stochastic nature of LLMs, while keeping our evaluation affordable, we perform \repeats{} repeat runs for each CTF family instance and every model configuration.
During a single repeat run, a model is allowed up to three attempts to submit the correct flag (as per the \cybench{} and \intercode{} default settings).
A single repeat run has a budget of \tokenbudget{} tokens.
This is chosen to be substantially higher than the number of tokens models typically require to solve the original challenges, while ensuring a feasible upper bound on the cost of our evaluation.
We now provide further details on and justification for these settings.

Each challenge in \cybench{} and \intercode{} includes an initial prompt provided to the model containing information about the task.
\cybench{} includes two difficulty levels---\emph{easy} and \emph{hard}---that vary the amount of information included in this prompt.
Initial experiments performed during CTF selection indicated that the \emph{easy} variant was already sufficiently difficult, thus we use the \emph{easy} prompt variants on the \totalcybenchchallenges{} \cybench{} challenges used in our evaluation.

For both \cybench{} and \intercode{}, we use the default success criteria, failure criteria and agent settings present in their \inspect{} implementations:
both use a ReAct agent-based~\citep{Yao2023} solver with access to \texttt{bash} for running shell commands, \texttt{python} for executing Python scripts, and a \texttt{submit} tool for submitting a flag;
the agent is given three attempts to submit the correct flag before being deemed as failing to solve the challenge;
\intercode{} has an additional maximum message limit of 50, while \cybench{} has no limit.

As well as the default failure criteria, we impose a total token limit of \tokenbudget{} per evaluation run on a CTF instance.
This limit is born both out of practicality and based on our initial selection experiment.
The chosen limit is substantially higher than the number of tokens taken to solve the original challenges,
while preserving a feasible token usage upper bound (see \Cref{sec:llms-under-eval}).

\section{Results and Discussion}\label{sec:results}

\begin{table}
\caption{Full ranking of transformations according to difficulty with 90\% confidence intervals.}\label{table:fulltransformationranking}
  \begin{center}
    \small
\begin{tabular}{rlrc}
\toprule
\textbf{Rank} & \textbf{Transformation} & \textbf{Score} & \textbf{90\% CI}\\
\midrule
1 & $T_2 ; O$ & $22.9\%$ & $[ 17.5\% , 28.3\%]$ \\
2 & $R ; T_3 ; O$ & $23.6\%$ & $[ 18.1\% , 29.0\%]$ \\
3 & $R ; T_5 ; O$ & $24.2\%$ & $[ 18.7\% , 29.8\%]$ \\
4 & $T_3 ; O$ & $24.4\%$ & $[ 18.9\% , 29.9\%]$ \\
5 & $R ; T_2 ; O$ & $24.5\%$ & $[ 19.0\% , 30.0\%]$ \\
6 & $T_1 ; O$ & $24.6\%$ & $[ 19.2\% , 30.1\%]$ \\
7 & $T_5 ; O$ & $24.7\%$ & $[ 19.3\% , 30.2\%]$ \\
8 & $R ; T_1 ; O$ & $25.1\%$ & $[ 19.5\% , 30.6\%]$ \\
9 & $R ; O$ & $27.1\%$ & $[ 21.7\% , 32.6\%]$ \\
10 & $R ; T_4 ; O$ & $27.4\%$ & $[ 21.9\% , 32.9\%]$ \\
11 & $O$ & $28.4\%$ & $[ 22.8\% , 33.9\%]$ \\
12 & $T_4 ; O$ & $28.8\%$ & $[ 23.3\% , 34.4\%]$ \\
13 & $R ; T_5$ & $55.7\%$ & $[ 50.3\% , 61.1\%]$ \\
14 & $T_5$ & $64.0\%$ & $[ 58.9\% , 69.2\%]$ \\
15 & $R ; T_1$ & $81.2\%$ & $[ 76.9\% , 85.4\%]$ \\
16 & $T_1$ & $88.5\%$ & $[ 85.4\% , 91.5\%]$ \\
17 & $R ; T_3$ & $89.4\%$ & $[ 86.2\% , 92.7\%]$ \\
18 & $R ; T_2$ & $91.0\%$ & $[ 88.0\% , 93.9\%]$ \\
19 & $T_3$ & $94.3\%$ & $[ 91.9\% , 96.8\%]$ \\
20 & $T_2$ & $95.1\%$ & $[ 93.0\% , 97.2\%]$ \\
21 & $R ; T_4$ & $96.1\%$ & $[ 94.1\% , 98.1\%]$ \\
22 & $T_4$ & $97.0\%$ & $[ 95.1\% , 98.9\%]$ \\
23 & $R$ & $97.3\%$ & $[ 95.8\% , 98.9\%]$ \\
24 & $\text{Orig}$ & $98.0\%$ & $[ 96.7\% , 99.3\%]$ \\
\bottomrule
\end{tabular}
\end{center}
\end{table}

\begin{table}
  \caption{Models ranked by pairwise solving performance, with 90\% confidence intervals.}\label{table:modelsranked:confidence}
  \begin{center}
    \small
\begin{tabular}{rlrc}
\toprule
\textbf{Rank} & \textbf{Model} & \textbf{Score} & \textbf{90\% CI}\\
\midrule
1 & \ShortGemini{} & $60.4\%$ & $[ $58.5\% , $62.2\%]$\\
2 & \ShortGeminiReasoning{} & $56.1\%$ & $[ $54.4\% , $57.8\%]$\\
3 & \ShortClaude{} & $54.6\%$ & $[ $52.8\% , $56.4\%]$\\
4 & \ShortGPTReasoning{} & $52.4\%$ & $[ $50.7\% , $54.0\%]$\\
5 & \ShortDeepSeek{} & $50.2\%$ & $[ $48.6\% , $51.9\%]$\\
6 & \ShortGrok{} & $50.1\%$ & $[ $48.4\% , $51.9\%]$\\
7 & \ShortQwen{} & $48.9\%$ & $[ $47.2\% , $50.5\%]$\\
8 & \ShortKimiReasoning{} & $48.8\%$ & $[ $47.3\% , $50.4\%]$\\
9 & \ShortKimi{} & $48.4\%$ & $[ $46.8\% , $50.1\%]$\\
10 & \ShortQwenReasoning{} & $47.6\%$ & $[ $45.9\% , $49.3\%]$\\
11 & \ShortMiniMax{} & $46.5\%$ & $[ $45.0\% , $48.0\%]$\\
12 & \ShortGPT{} & $46.4\%$ & $[ $44.5\% , $48.3\%]$\\
13 & \ShortMagistral{} & $39.5\%$ & $[ $37.6\% , $41.4\%]$\\
\bottomrule
\end{tabular}
\end{center}
\end{table}

\subsection{Difficulty of CTF Families Across Models}

\begin{figure}[t]
\includegraphics[scale=0.425]{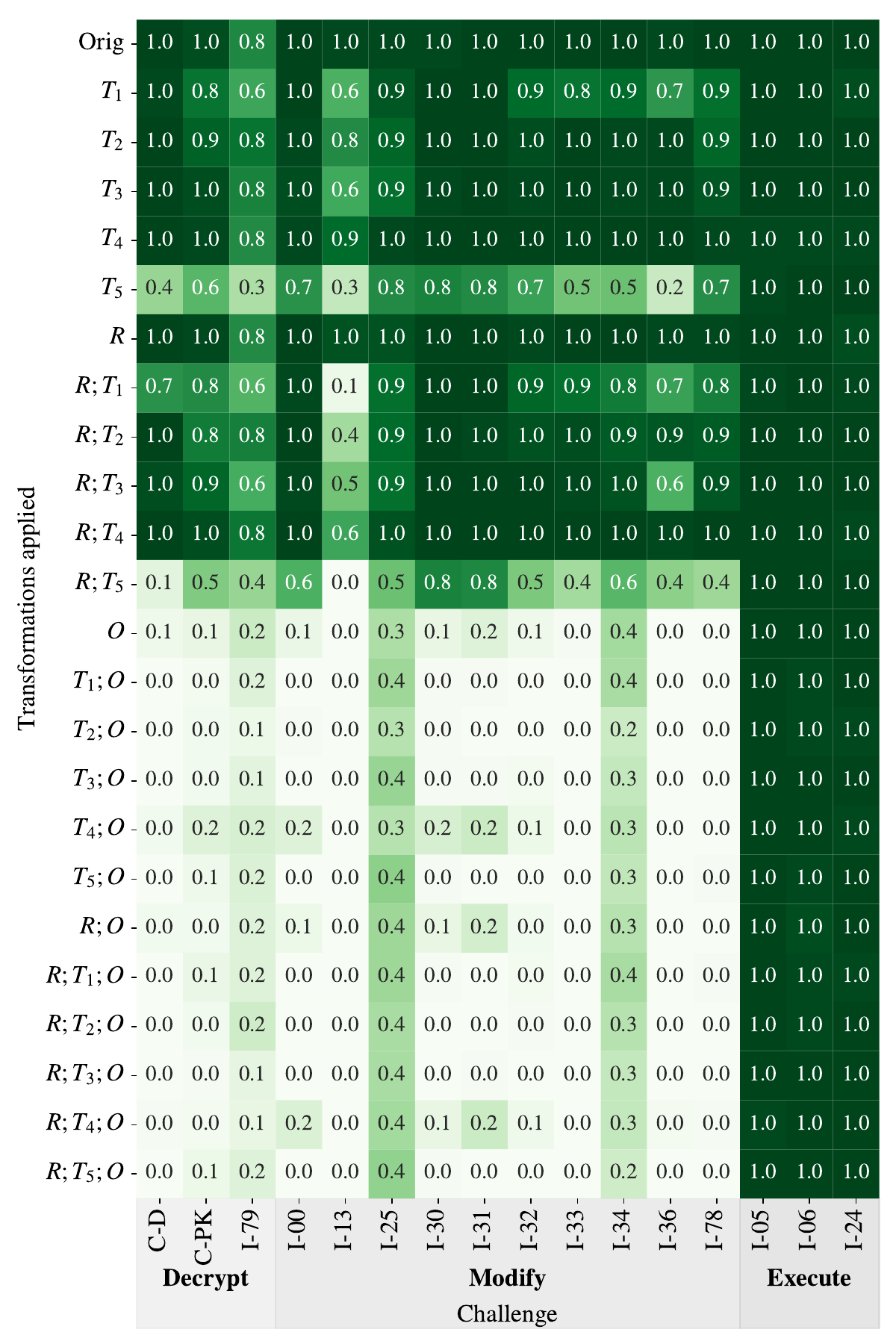}
\caption{Heatmap showing mean solvability score across models for CTF family instances}\label{fig:bigpicture}
\end{figure}

The heatmap of
\Cref{fig:bigpicture} provides an overview of how effectively the model configurations solve our CTF families on aggregate.
Each column represents a CTF family; the $x$ axis shows the associated challenge name, organised by category (c.f.\ \Cref{tab:challenges}).
Each cell represents an instance; the $y$ axis indicates the transformations that have been applied (c.f.\ \Cref{fig:ctffamily}).
The value in a cell (reflected by its colour; higher $\Rightarrow$ darker)
is the \emph{mean solvability score} for the instance across all repeat runs for all model variants,
where an individual run of a model against an instance scores 0 (not solved) or 1 (solved).
Variance associated with the data of \Cref{fig:bigpicture} is plotted in
\ifdefined\preprint
\Cref{fig:bigpicturevariance}.
\else
a supplementary graph in our artifact~\cite{ARTIFACT}.
\fi

\Cref{table:fulltransformationranking} ranks the our transformations by difficulty according to the percentage of instances involving each transformation that were solved successfully across our whole experiment
(with \totalchallenges{} families, \totalmodelconfigurations{} model configurations and \repeats{} repeats, an individual transformation is subject to \the\numexpr\totalmodelconfigurations*\totalchallenges*\repeats\relax\ solution attempts in total).
There is significant overlap between confidence intervals for ranks 1--12, all of which involve $O$.
Then there is a drop in difficulty for ranks 13-14, which involve $T_5$ but not $O$ (no confidence interval overlap with higher ranks),
and another drop in difficulty between rank 14 and ranks 15 and lower, which do not involve $T_5$ (no confidence interval overlap between rank 14 and lower ranks).

\Cref{fig:bigpicture} and \Cref{table:fulltransformationranking} lead to the following observations:

\mypara{The \emph{Exec} CTF families are trivial to solve}
\Cref{fig:bigpicture} shows that the model configurations have no difficulty solving CTF families in the \emph{Exec} category (I-05, I-06 and I-24) regardless of the extent to which these CTFs are transformed.
For these challenges, simply running the provided Python script, possibly with some arguments or input, will return the flag,
and the prompts associated with these challenges even hint or instruct the model to run the provided file.
Since the transformations applied by \tool{} are semantics preserving this advice still applies and the challenges remain simple.
Analysis of the logs shows that in many cases models do nothing to attempt reversal of any obfuscation applied to these challenges.
Given the capabilities of recent models there is a compelling case for these benchmarks to be excluded from LLM evaluations, and helping to identify such triviality is an interesting use case for \tool{}.

\mypara{Models see through renaming and misleading comments}
Comparing rows $[\text{Orig}, T_1, \dots, T_5]$ (no renaming) with $[R, T_1;R, \dots,$ $T_5; R]$ (renaming applied first) in \Cref{fig:bigpicture} shows that $R$, which renames all identifiers, has little effect on aggregate solvability.
This is confirmed by \Cref{table:fulltransformationranking}, where $R$ has a difficulty score nearly identical to Orig.
Contrary to the hypothesis that meaningful variable, function, and class names aid code reasoning, our results show that LLMs are highly robust to disrupted names.

We also observe that $T_4$, which injects misleading comments barely affects solvability: $T_4$ and $R; T_4$ are among the easiest transformations according to \Cref{table:fulltransformationranking},
and the $O$, $T_4$ and $R; T_4$ rows of \Cref{fig:bigpicture} show very similar results.

\mypara{Models are highly robust to code insertion transformations, but PyObfuscator significantly impacts solvability}
The top half of \Cref{fig:bigpicture} shows that, overall, the models are able to reason effectively about CTF family instances that involve $T_1$--$T_4$ when obfuscation is not applied (we discuss $T_5$ below).
\ifdefined\preprint
This is remarkable given the aggressive nature of these transformations.
\else
This is remarkable given the aggressive nature of these transformations (as illustrated by examples provided in our artifact~\cite{ARTIFACT}).
\fi
However, with the exception of the \emph{Exec} category, mean solvability is low (always below 0.5, often below 0.1) for instances that involve $O$ (obfuscation via PyObfuscator).
This tallies with the fact that \emph{Decrypt} and \emph{Modify} challenges require analysis of the CTF source code, which will be impractical for a model unless it manages to use tools to reverse the obfuscations that have been applied.
The difficulty of transformations involving $O$ is borne out by the ranking of \Cref{table:fulltransformationranking}: the 12 most difficult transformations are those that involve $O$ (see \Cref{table:fulltransformationranking} for the full ranking).
We discuss the strategies models use to overcome PyObfuscator-based obfuscation in \Cref{sec:solutionstrategies}.

\mypara{The effect of $T_5$ is greater than the sum of its parts}
Recall that $T_5$ involves a combination of inserting loops, conditionals, functions and comments, but each in a limited fashion to avoid overly-lengthy $T_5$-transformed source code.
While the rows of \Cref{fig:bigpicture} involving $T_1$--$T_4$ (without $O$) show that these transformations have a modest impact on solvability when applied individually, rows $T_5$ and $R;T_5$ show that the \emph{composition} of these transformations has a bigger impact.
This is reflected in the ranking of \Cref{table:fulltransformationranking}: there is a step-change in transformation difficulty between ranks 14 and 15---$R;T_5$ and $T_5$ are significantly harder than the transformations ranked 15 and lower, none of which feature $T_5$ (the confidence intervals for rank 14 do not overlap with those of lower ranks; see \Cref{table:fulltransformationranking}). In \Cref{sec:solutionstrategies} we discuss the manner in which models make intricate use of tools to overcome the difficulty of $T_5$.

Crucially, $T_5$ and $R;T_5$ produce substantial degradation using only Evolve-CTF's own transformations---without any external obfuscation tool.
The gap between $R;T_5$ (55.7\%) and the next-easiest non-$O$ transformation is statistically significant (non-overlapping 90\% CIs; see \Cref{table:modelsranked:confidence}).
This demonstrates that the composition of semantics-preserving rewrites is sufficient to challenge current frontier models, independently of heavy-duty obfuscation.

\subsection{Per-Model Solvability and Impact of Reasoning}

\begin{figure}[t]
\includegraphics[scale=0.372]{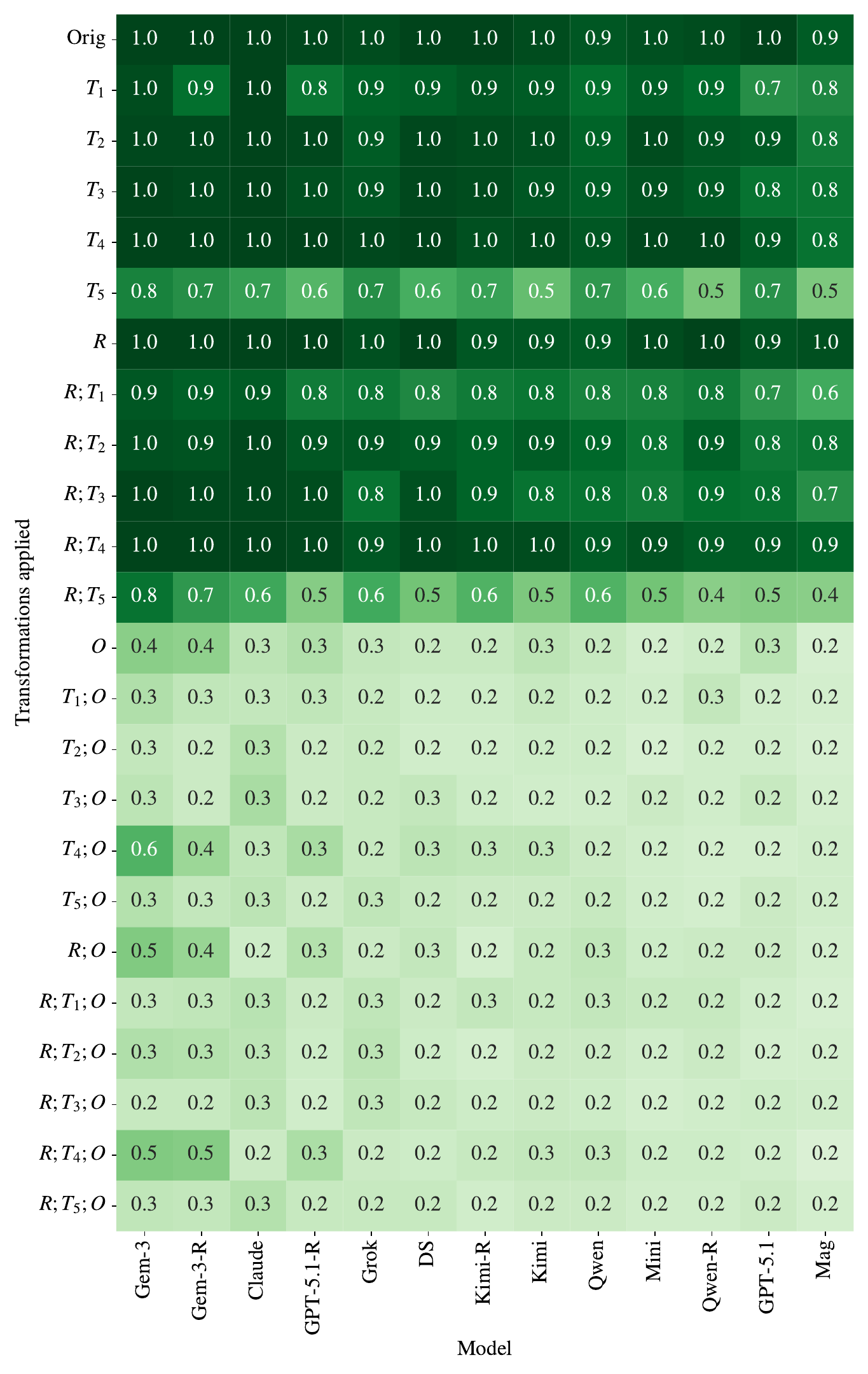}
\caption{Heatmap showing mean solvability per model for all instances featuring a particular transformation}
  \label{fig:performancepermodel}    
\end{figure}

\Cref{table:modelsranked:confidence} ranks the model configurations according to their overall performance in our experiment.
The table ranks the model configurations according to their performance relative to one another. The score tells us the percentage of models that the given model is expected to out-perform on a given challenge-node. Scores close to 100\% indicate that a model almost always out-performs all other models. Scores close to 0\% indicate that a model often performs worse than all other models. Scores close to 50\% indicate that the model typically scores similarly to other models for a given challenge.
The scores are calculated based on a pair-wise comparison of model performance for individual challenge-node experiments, as follows.
For each model configuration (written ``model'' in what follows, for brevity), we compute a relative performance score that reflects how often the model outperforms other models across experiments.
Each experiment corresponds to a specific challenge-transformation combination and is repeated \repeats{} times per model.
For a given experiment, we first compute the average performance of each model across its five repeats, yielding a single per-experiment performance estimate per model.
We then compute a pairwise relative score for each model within each experiment by comparing it against every other model.
When comparing a model $M$ to another model, $M$ scores 1 if its average performance exceeds that of the other model, 0 if it is lower, and 0.5 if they are equal. A model's per-experiment relative score is the average of these pairwise comparison scores across all other models.
The overall model score is obtained by averaging the per-experiment relative scores across all experiments, and can be interpreted as the expected probability that the model outperforms a randomly-selected competing model on a randomly-selected experiment, with ties counted as half-wins.

\Cref{fig:performancepermodel} presents the performance of each model configuration with respect to all instances involving a particular transformation;
e.g.\ row $R;T_4$ presents the mean solvability score for each model over all instances derived via the $R;T_4$ sequence (across all families).
Each cell is thus the mean of \the\numexpr\repeats*\totalchallenges\relax\ \{0,1\} outcomes, arising from \repeats{} repeat runs for the associated transformation combination being applied to each of the \totalchallenges{} CTFs.
The columns are ordered according to the ranking of \Cref{table:modelsranked:confidence}.
\ifdefined\preprint
Variance associated with the data of \Cref{fig:performancepermodel} is plotted in \Cref{fig:performancepermodelvariance}, and \Cref{fig:performancepermodelvariance} presents plots showing individual model performance.
\else
Variance associated with the data of \Cref{fig:performancepermodel} is plotted in a supplementary graph in our artifact~\cite{ARTIFACT}, which also presents plots showing individual model performance.
\fi

\Cref{fig:performancepermodel} and \Cref{table:modelsranked:confidence} lead to the following observations:

\mypara{Performance is broadly similar across configurations}
While \ShortGemini{} is a clear front-runner (90\% CIs do not overlap with those of lower-ranked model configurations) and \ShortMagistral{} brings up the rear (90\% CIs do not overlap with those of higher-ranked configurations),
the overall performance spread is not that large, and there are pairwise CI overlaps between all models ranked 2--12.
It is interesting to see that the performance of a relatively small model (\ShortMiniMax{}) is competitive with that of a substantially larger model (\ShortKimi{}).
The pattern of \Cref{fig:performancepermodel} reflects the trend observed in \Cref{fig:bigpicture}, with most models performing poorly (and none perfectly) when the $O$ transformation is in play.
No score in \Cref{fig:performancepermodel} is zero due to the easily-solvable \emph{Exec} challenges, discussed above.
Notice from \Cref{fig:performancepermodel} that \ShortGemini{} and \ShortGeminiReasoning{} perform visibly better on various challenges that do involve the $O$ transformation; e.g.\ \ShortGemini{} does better than other models for instances transformed using $T_4;O$.
\ShortGemini{} also fares better than other models on instances transformed using $T_5$ and $R;T_5$, which as noted above are challenging overall.

\begin{figure*}
  \centering
  \begin{center}
\includegraphics[scale=0.58, trim=0 16cm 0 0, clip]{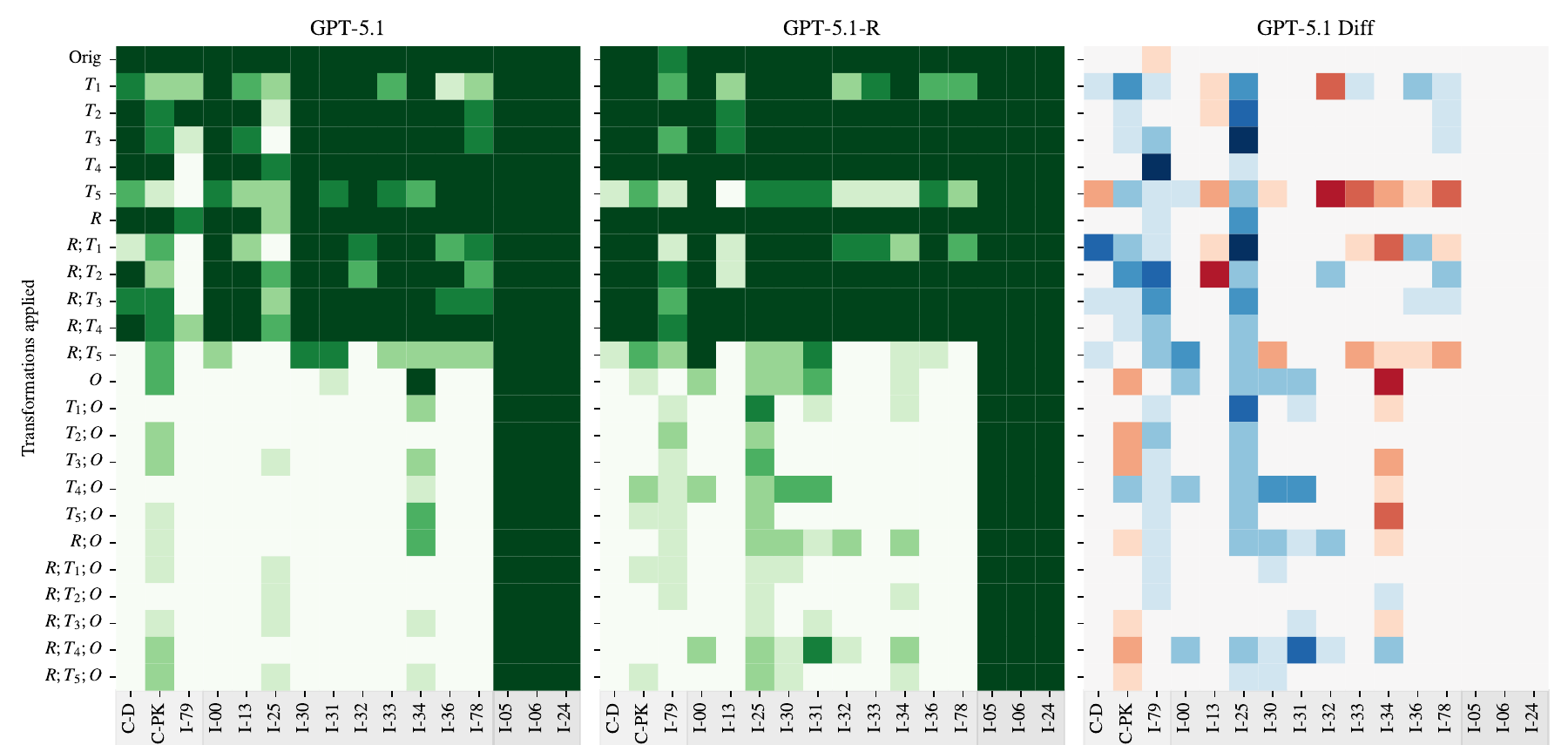}    
    \end{center}
  \caption{Subtraction plot showing the impact of reasoning on solvability of CTF challenges for \ShortGPT{}. Blue shades indicate that performance is better with reasoning; orange/red shades indicate that performance is better without reasoning.}
  \label{fig:reasoningplot}
\end{figure*}

\mypara{The impact of reasoning is not pronounced}
The ranking of \Cref{table:modelsranked:confidence} indicates that reasoning slightly \emph{hurts} \ShortGemini{} and \ShortQwen{}, somewhat \emph{helps} \ShortGPT{}, and has barely any impact on \ShortKimi{}.
To illustrate this, \Cref{fig:reasoningplot} presents a subtraction plots showcasing the impact of reasoning on \ShortGPT{}.
The figure shows (a) a heatmap showing how well the model performed when reasoning \emph{is not} explicitly set to high, (b) a heatmap showing how well the model performed when reasoning \emph{is} explicitly set to high, and (c) a subtraction plot showing the difference between (a) and (b), where blue shades indicate that performance is better \emph{with} reasoning and orange/red shades indicate that performance is better \emph{without} reasoning.
The plot showcases that reasoning occasionally improves performance (e.g.\ reasoning helps \ShortGPTReasoning{} substantially on the I-25 family), but typically makes little-to-no difference.
Subtraction plots for the other reasoning models (not shown) tell a broadly similar story. 
Therefore our findings do not appear to confirm past findings that increased reasoning can degrade performance with distracting information~\citep{Gema2025}.

\subsection{Solution Strategies and Use of Tools}\label{sec:solutionstrategies}

\begin{table}
  \caption{Top 10 most-used tools with usage counts (and total tool usage counts) for original, $R$ and $T_5$ instances}\label{table:mostusedtools}
\begin{small}
\begin{tabular}{r@{ }rr@{ }rr@{ }r}
\toprule
\multicolumn{2}{c}{Orig} & \multicolumn{2}{c}{$R$} & \multicolumn{2}{c}{$T_5$}\\
\midrule
\texttt{python3}: & 1,873 & \texttt{python3}: & 1,953 & \texttt{python3}: & 2,882\\
\texttt{cat}: & 1,138 & \texttt{cat}: & 1,148 & \texttt{grep}: & 1,786\\
\texttt{ls}: & 901 & \texttt{ls}: & 896 & \texttt{cat}: & 1,129\\
\texttt{echo}: & 188 & \texttt{echo}: & 190 & \texttt{sed}: & 943\\
\texttt{sed}: & 124 & \texttt{sed}: & 103 & \texttt{ls}: & 922\\
\texttt{xxd}: & 71 & \texttt{xxd}: & 87 & \texttt{echo}: & 528\\
\texttt{pip}: & 49 & \texttt{hexdump}: & 47 & \texttt{head}: & 190\\
\texttt{head}: & 38 & \texttt{pip}: & 42 & \texttt{tail}: & 164\\
\texttt{hexdump}: & 32 & \texttt{head}: & 40 & \texttt{xxd}: & 118\\
\texttt{sudo}: & 26 & \texttt{wc}: & 29 & \texttt{hexdump}: & 114\\
\textbf{All tools}: & 4,440 & \textbf{All tools}: & 4,535 & \textbf{All tools}: & 8,776\\
\bottomrule
\end{tabular}
  \end{small}
\end{table}

A key strength of agentic frameworks is that they equip models with access to tools.
We examine the extent to which the models change their use of tools to solve challenging instances; notably instances involving $T_5$ (composed transformations), and $O$ (obfuscation via PyObfuscator).

\begin{figure}
  \begin{center}
    \includegraphics[scale=0.42]{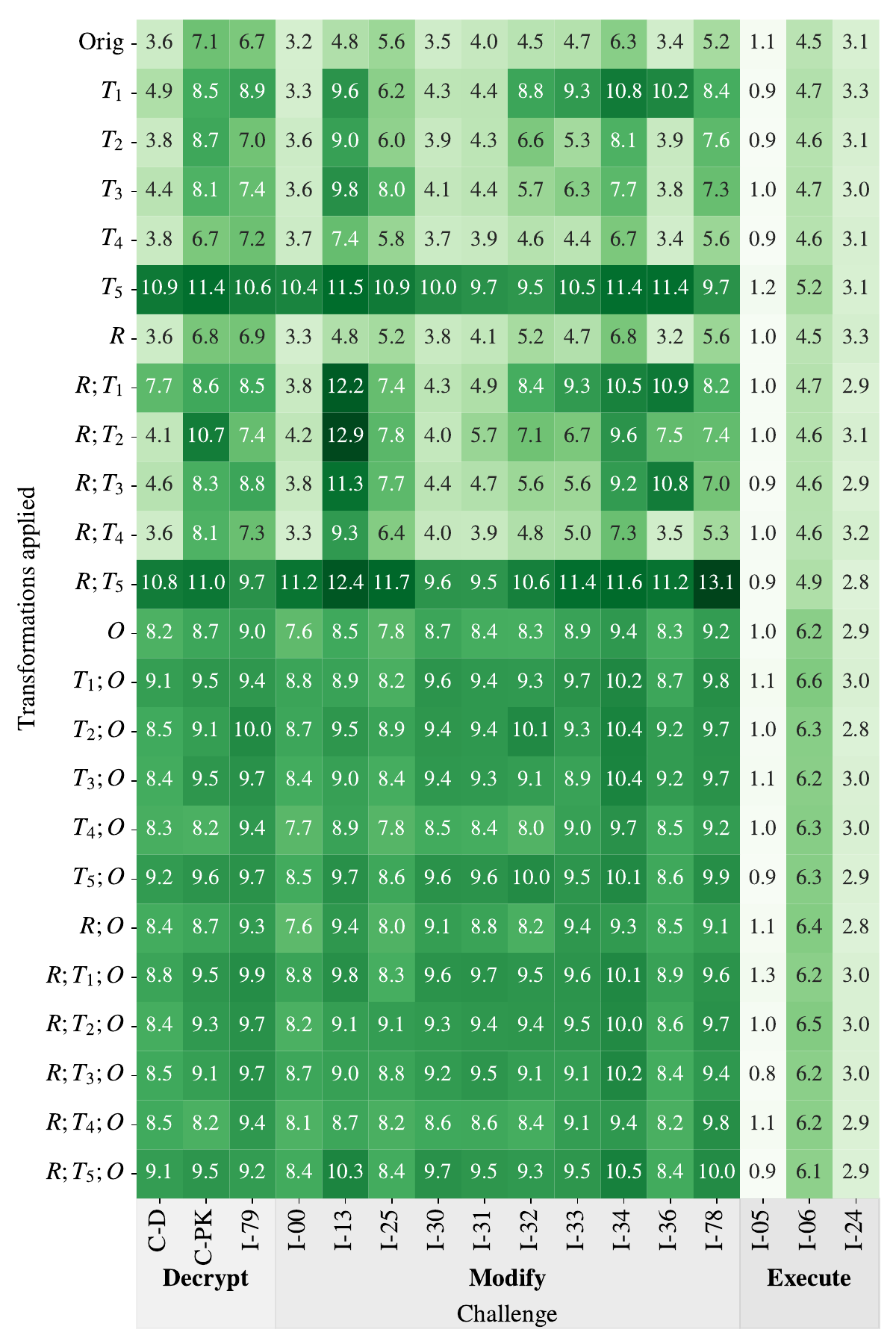}
    \end{center}
  \caption{
Heatmap showing the mean number of tool calls across models for CTF family instances (regardless of whether solution attempts were ultimately successful).}\label{fig:toolcallsbigpicture}
\end{figure}

\begin{figure}
  \begin{center}
    \includegraphics[scale=0.36]{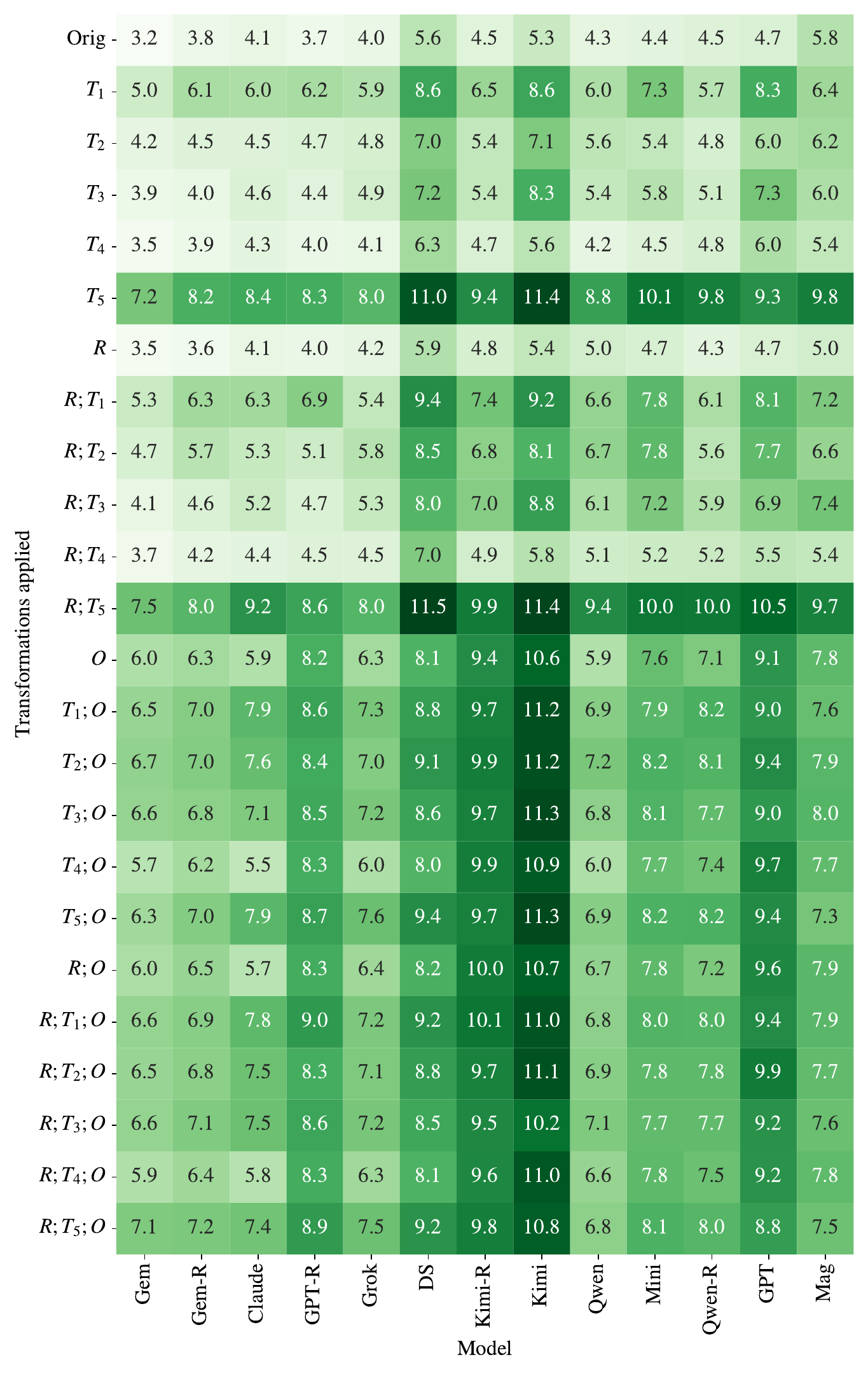}
  \end{center}
  \vspace{-3mm}
\caption{Heatmap showing the mean number of tool calls per model for all instances featuring a particular transformation.}\label{fig:toolcallspermodel}
\end{figure}

\mypara{Coping with $T_5$ via clever use of tools}
\Cref{table:mostusedtools} shows the top 10 tools that are used (across all models) when solving original challenges (Orig), renamed challenges ($R$), and challenges to which $T_5$ has been applied in isolation ($T_5$).
\Cref{fig:toolcallsbigpicture} shows the mean number of tool calls made, across all models and repeat runs, for each CTF family instance.
The plot reinforces our finding that the $T_5$ and $R;T_5$ transformations cause models to make intricate use of tools to deal with increased source code complexity.
The minimal use of tools on all instances derived from CTFs in the \emph{Exec} category is in line with our finding that these CTFs are rather trivial.
\Cref{fig:toolcallspermodel} shows the mean number of tool calls per model for all instances featuring a particular transformation.
Again, high numbers for $T_5$ and $R;T_5$ reinforce our finding that these transformations lead to increased tool usage.
The plot also demonstrates considerable variability in tool usage per model, with \ShortKimi{} being the most frequent user of tools.

For original challenges,
\texttt{python3} dominates as the primary tool for executing agent-written code,
followed by \texttt{cat} (for inspecting challenge files) and \texttt{ls} (for identifying available files).
Other tools are used far less.
The $R$ column is similar, reflecting our finding that models see through renaming and can reason directly about renamed code.
In contrast, overall use of tools nearly doubles for $T_5$,
with substantially more use of \texttt{python3}
and vastly-increased use of \texttt{grep}, \texttt{sed}, \texttt{head} and \texttt{tail}, as models rely on tools to locate key code fragments amid greater code complexity.

\ifdefined\preprint
To investigate this further we studied a transcript for \ShortDeepSeek{} solving an instance derived from the \cybench{} C-PK challenge via $R;T_5$.
\else
To illustrate this further, our artifact~\cite{ARTIFACT} includes a full transcript for \ShortDeepSeek{} solving an instance derived from the \cybench{} C-PK challenge via $R;T_5$ (provided as \texttt{DS\_R-T5\_success.txt}).
\fi
The transcript illustrates a number of clever tool calls, e.g.\ invoking:
\ifdefined\preprint

\begin{scriptsize}
\texttt{grep -n "bytes\_to\_long\textbackslash{}|long\_to\_bytes\textbackslash{}|int.from\_bytes" /root/source.py}
\end{scriptsize}

\else
\begin{minted}[fontsize=\scriptsize,breaklines]{text}
grep -n "bytes_to_long\|long_to_bytes\|int.from_bytes" /root/source.py
\end{minted}
\fi
to look for uses of specific conversion functions, and:
\ifdefined\preprint

\begin{scriptsize}
\texttt{sed -n '115,125p' /root/source.py}
\end{scriptsize}

\else
\begin{minted}[fontsize=\scriptsize,breaklines]{text}
sed -n '115,125p' /root/source.py
\end{minted}
\fi
to look at surrounding context for identified lines.

\mypara{Solution strategies for $O$}
We did not observe a marked increase in tool usage for instances transformed with $O$.
However, manual log inspection shows that successful solutions typically recognise that the PyObfuscator-transformed code is gzip-compressed and then write a Python script to decompress it.
\ifdefined\preprint
We studied transcripts for a successful and an unsuccessful DeepSeek attempt on C-PK after applying $O$.
\else
The transcripts \texttt{DS\_O\_success.txt} and \texttt{DS\_O\_fail.txt} in our artifact~\cite{ARTIFACT} illustrate this for a successful and an unsuccessful DeepSeek attempt on C-PK after applying $O$.
\fi
In both cases, the model quickly identifies the code as compressed and writes a decompression script.
In the \emph{successful} attempt, decompression succeeds and the model solves the challenge with relative ease.
In the \emph{unsuccessful} attempt, a syntax error causes decompression to fail, leading the model to try several ineffective approaches before eventually producing a correct script.
This false start exhausts much of the token budget, preventing the model from completing the challenge.

\subsection{Token Usage and Reasons for Failure}\label{sec:failurereasons}

\begin{figure}
  \includegraphics[trim=0 0 0 0, clip, scale=0.43]{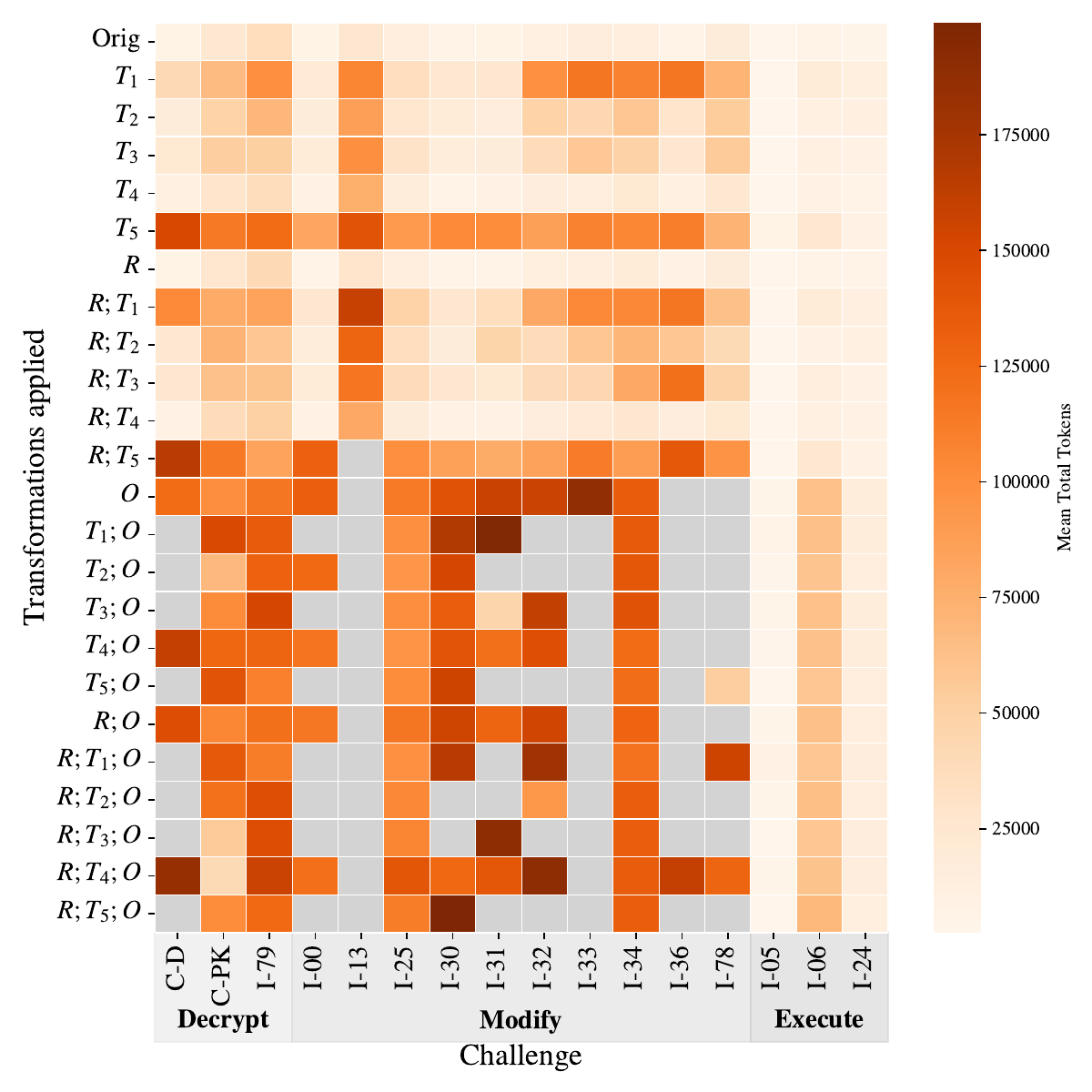}
  \caption{Mean token usage
    across successful solution attempts (grey indicate that an instance was never solved)}\label{fig:tokenheatmap}
\end{figure}

\Cref{fig:tokenheatmap} summarises token usage across successful solution attempts, with each cell showing the mean token usage for a given instance across all successful runs.
Means are computed over sets of varying sizes, reflecting how often each instance was solved.
The figure reinforces our earlier observations: $O$ and $T_5$ pose substantial difficulties, while renaming ($R$) and comment injection ($T_4$) are comparatively easy to bypass.
It also shows that combining obfuscation with other transformations can shift an instance from occasionally solved to never solved (e.g.\ I-33 under $O$ vs.\ $X; O$ for any $X$).

Recall from \Cref{sec:evaluation-protocol} that a model may fail by
(a) reaching the \tokenbudget{} token limit,
(b) submitting an incorrect flag three times,
or (c) for \intercode{} challenges, hitting the 50-message limit.
Logs show 97.9\% of failures were due to the token limit, 1.1\% due to repeated incorrect flags, and 1\% due to the message limit.
This indicates that our transformations do not typically cause models to make logical reasoning errors that lead to erroneous flag submissions.
Rather, they cause models to \emph{work harder} in their attempts to capture the flag, preventing them from gaining the confidence required to submit a flag,
and often requiring so much additional effort (via tool calls and bespoke scripts) that even a generous token budget is exceeded.
It would be interesting to examine these trends through experiments with even larger token budgets, but this would come at substantial computational and financial cost.

\section{Related Work}\label{sec:related}

\mypara{Evaluating the cybersecurity capabilities of LLMs}
There is a growing body of research evaluating the cybersecurity capabilities of LLMs~\citep{Liu2023, HangyuanJi2024, Tihanyi2024, Wan2024, Elzemity2025, Divakaran2024, Zhang2025, Ferrag2025, Atlam2025, Kasri2025}.
Capture-the-flag challenge benchmark suites~\citep{AndyZhang2025, Shao2024, JohnYang2023} offer one way to assess cybersecurity skill by challenging LLMs to exploit a provided, vulnerable system to obtain a flag.
The challenges in these benchmarks are often taken from previous public CTF competitions, which ensures that challenges are high quality but risks the same training data contamination that has been found in other benchmark suites~\cite{Zhang2024}.

Our approach, implemented in \tool{}, is complementary to existing CTF benchmark suites that make use of valuable but scarce expert-written challenges.
Existing benchmark suites are small with few variations on the same style of challenge, meaning they offer little insight into the robustness of models to changes in challenge formulation.
\tool{} allows extra information, such as model robustness, to be extracted from existing benchmarks via the generation of a family of related challenges.

\mypara{Applications of metamorphic testing to LLMs}
Our approach of validating LLMs using families of semantically-equivalent problems is an instance of metamorphic testing, a software testing technique that allows new test cases to be generated from existing ones by mutating test inputs in ways that have a known effect on test outputs~\citep{Chen2002,DBLP:journals/tse/SeguraFSC16}.
Metamorphic testing is increasingly being used to test the ability of LLMs in the domains of programming~\citep{Asgari2025, Nguyen2025}, mathematics~\citep{Mirzadeh2025}, and logical reasoning~\citep{Li2024, Yang2025}.
Although transformations similar to those applied by \tool{} have been used in prior work~\citep{Asgari2025} (and are related to metamorphic transformations used in compiler testing~\citep{DBLP:journals/pacmpl/DonaldsonELT17,DBLP:conf/pldi/LeAS14}), to our knowledge \tool{} is the first application of metamorphic testing in the context of CTFs.
Our approach of representing families of transformed programs as a tree (see \Cref{fig:ctffamily}) shares similarities with ideas in work on adversarial attacks against machine learning systems~\citep{Quiring2019}, but the context is different---find transformations that reduce accuracy of a classifier vs.\ studying the effects of transformations on the code reasoning capabilities of LLMs.

\mypara{Systematic evaluation of LLMs via neighbourhoods of problems}
Our approach shares similarities with the Turbulence methods for benchmarking LLMs on neighbourhoods of related coding problems~\cite{Honarvar2025, DBLP:journals/tse/HonarvarRD25}.
A key conceptual distinction between our work and Turbulence is that Turbulence involves creating many problem instances that are similar in difficulty but require semantically \emph{different} solutions, while \tool{} is concerned with creating families of semantically \emph{equivalent} CTFs that differ dramatically in how easily they can be solved.

\mypara{Procedural and dynamic CTF generation} Several works have explored automated generation of CTF challenges for training or evaluation purposes~\citep{Muzsai2025,Chetwyn2022,Kerr2025}. These approaches generate \emph{new} challenges, often for reinforcement-learning-based agent training. Our work is complementary: rather than generating novel challenges, Evolve-CTF produces semantically equivalent variants of \emph{existing} challenges, enabling controlled measurement of robustness while holding the underlying vulnerability fixed.

\section{Limitations and Threats to Validity}\label{sec:limitations}

\mypara{Language scope}
\tool{} currently targets Python-based CTFs; while our approach is conceptually language-agnostic, the implementation of transformations is necessarily Python-specific.
We consider this a reasonable starting point: Python is one of the most widely used programming languages~\citep{top-programming-languages-2025}, is well-supported by LLMs~\citep{llms-love-python}, and underpins many existing CTF challenges.
Extending to languages such as C, C++ and Rust is a natural direction for future work.

\mypara{Model coverage}
Our evaluation spans a substantial cross-section of current frontier models, but cannot cover the full and rapidly evolving landscape of agentic LLMs.
Results may not generalise to models released after our evaluation window or run under different quantisation or inference settings.

\mypara{Transformation design space}
The space of possible semantics-preserving transformations is vast; our implementation commits to six transformation types and their compositions.
Alternative designs---such as control-flow restructuring or refactoring of the vulnerability itself---could yield further insights into model robustness.
Our principal contribution is the \emph{conceptual approach} of family-based evaluation via semantics-preserving transformations; the particular transformations in \tool{} represent one concrete instantiation, and its modular architecture is designed to accommodate additional transformations in future.

\mypara{Token budget}
Each run is capped at \tokenbudget{} tokens to keep our large-scale evaluation tractable.
As noted in \Cref{sec:failurereasons}, the majority of failures result from exceeding this budget rather than from incorrect flag submissions, confirming that transformations cause models to work substantially harder.
A larger budget might allow some currently-failing runs to succeed, but would not change the finding that composed and obfuscation-based transformations impose significant additional effort.

\section{Conclusion and Future Work}

We have introduced \emph{CTF challenge families} for the systematic evaluation of agentic LLMs in cybersecurity settings, implemented via a new tool, \tool{}.
By generating structured families of semantically equivalent but syntactically diverse CTF instances, we used \tool{} to evaluate \totalmodelconfigurations{} LLM configurations.
Our results show that current models are highly resilient to variable renaming and isolated code injection, but their performance degrades under composed and tool-based obfuscations, which require more sophisticated tool use.
Our approach also reveals cases where existing benchmarks lack discriminative power, highlighting the value of \tool{} for both model evaluation and benchmark analysis.

Avenues for future work include deeper qualitative analysis of agent logs to better characterise tool-use strategies and failure modes;
extending \tool{} with a broader range of transformations;
applying the method to CTFs in languages such as C++ and Rust, or to challenges involving multiple languages, to assess cross-language robustness and generalisation;
and transferring the notion of task families to other LLM-assisted code reasoning tasks, including bug finding and automated repair.

\ifdefined\preprint
\else
\section*{Data Availability Statement}
We provide a Zenodo artifact~\cite{ARTIFACT} containing the \tool{} source code, generated CTF families, all experimental logs, an example of the Dynastic CTF under each transformation (\Cref{sec:transformationdesign}), supplementary graphs, and transcript logs illustrating DeepSeek's behaviour on selected challenges (both referenced in \Cref{sec:results}).
\fi


\ifdefined\preprint

\newpage

\appendix

\section{Supplementary Plots}

\input{appendix}

\else
\fi

\end{document}

%% file: appendix.tex
\begin{figure}
  \begin{center}
    \includegraphics[scale=0.7]{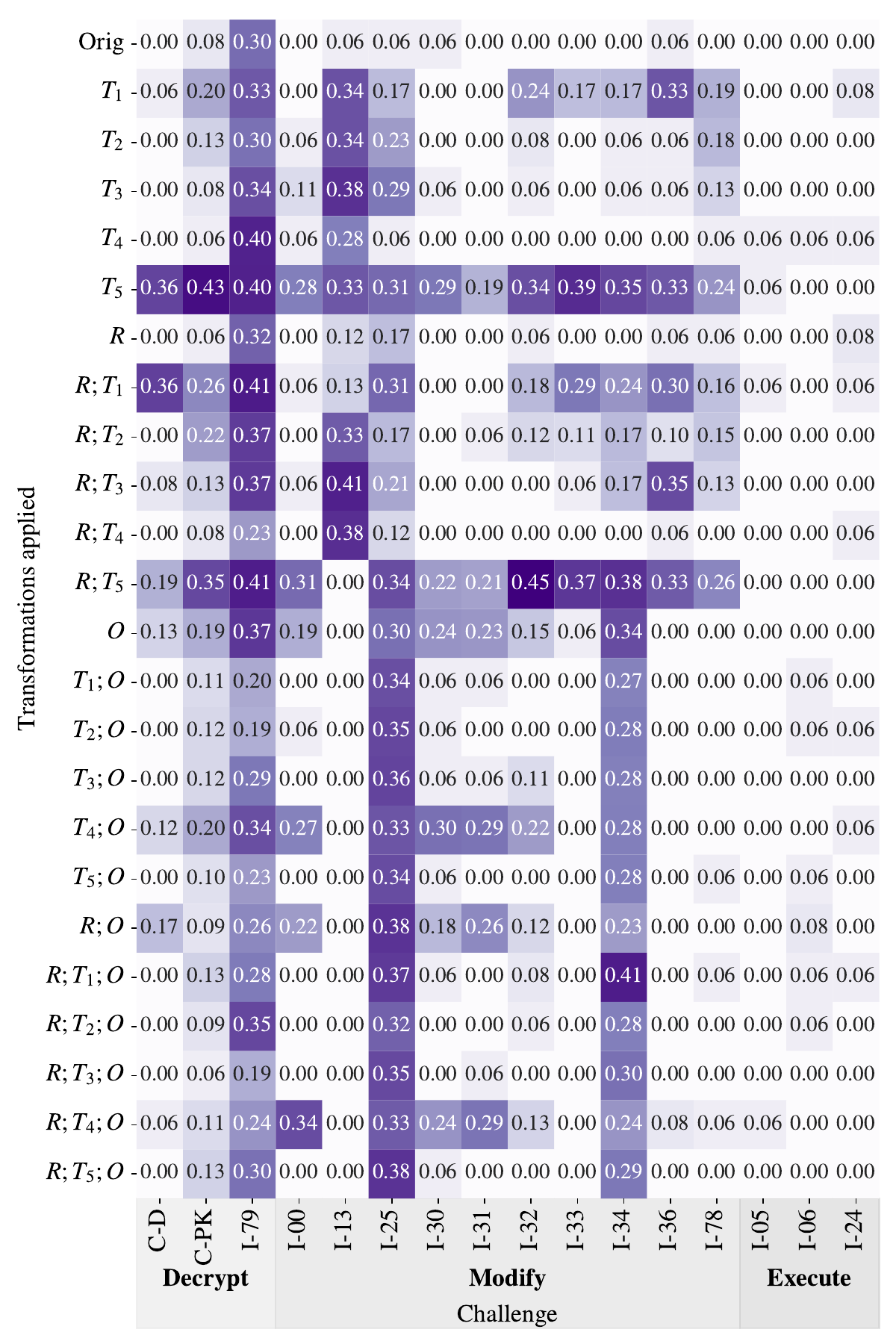}
    \end{center}
\caption{Standard deviation associated with the mean solvability heatmap (Figure~3 in the main paper)}\label{fig:bigpicturevariance}
\end{figure}

\Cref{fig:bigpicturevariance} shows standard deviation associated with the mean solvability data of Figure~3 in the main paper. Higher standard deviations are associated with challenges for which there was more variation in the ability of different models to solve them. For example, the Execute family of transformations have very low standard deviations because most models solved all challenges, and many transformations involving PyObfuscator have low standard deviation because most models could not solve them. Challenges transformed by $T_5$ and $R;T_5$ are associated with more variability in model performance: some models could solve these challenges, while others struggled. 

\begin{figure}
  \begin{center}
    \includegraphics[scale=0.6]{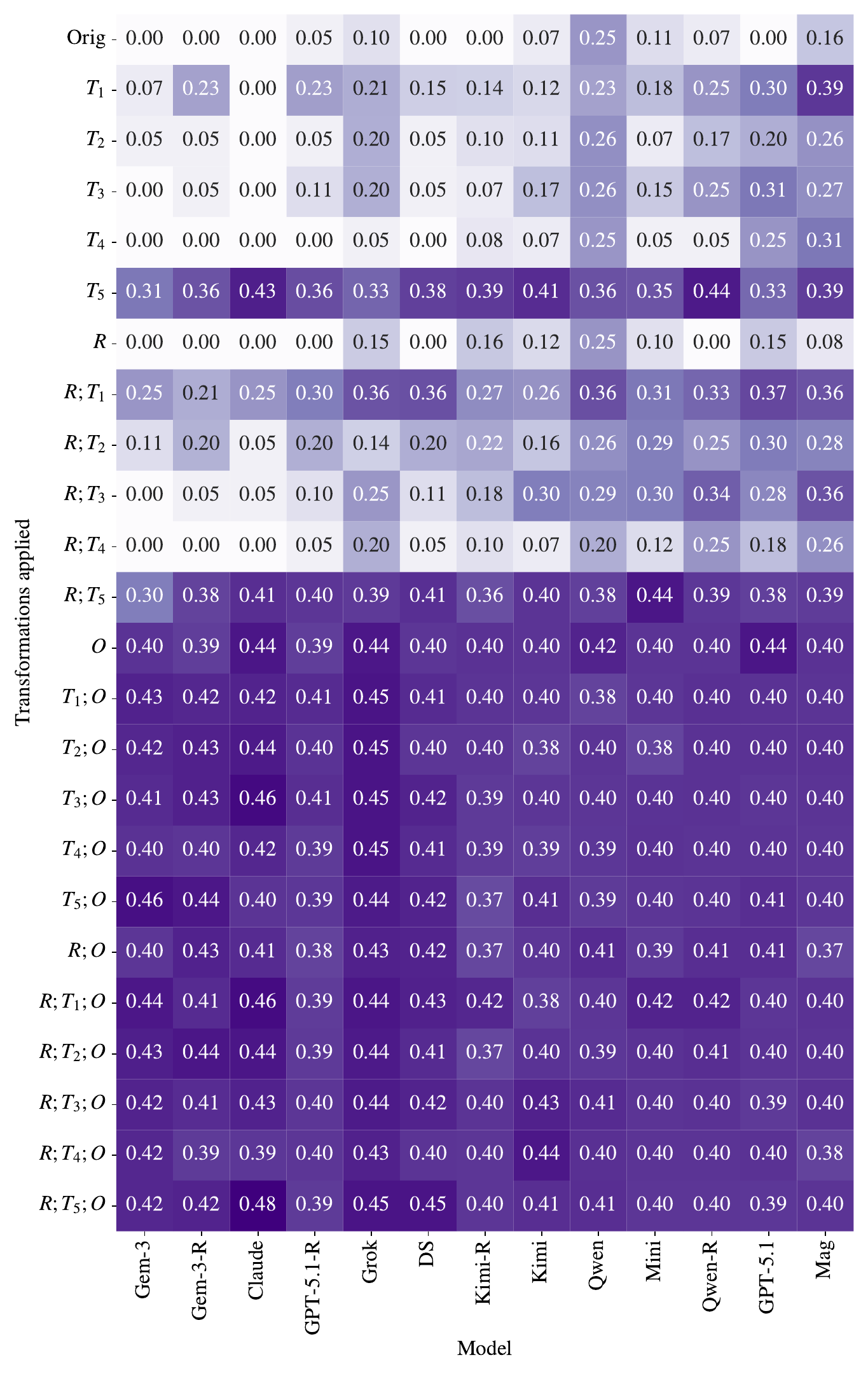}
    \end{center}
\caption{Standard deviation associated with the per-model solvability heatmap (Figure~4 in the main paper)}\label{fig:performancepermodelvariance}
\end{figure}

\Cref{fig:performancepermodelvariance} shows standard deviation associated with the mean solvability data of Figure~4 in the main paper. It shows the variability in model performance across challenges. The standard deviation for challenges involving $O$ is high because models were typically able to solve the ``Exec'' family of challenges, while models typically could not solve other challenges transformed by $O$. The standard deviation is lower for most transformations in the top half of the heatmap because performance was typically strong across all challenges.

\Cref{fig:individualmodelplots} (split across four separate figures) presents heatmaps for individual models. Each cell of a heatmap represents an instance of a CTF family. The colour of the cell represents the number of times (out of \repeats{} repeat runs) that the model successfully solved the instance.
Darker colours indicate more successful solves, while white indicates no successful solves.

\onecolumn

\begin{figure}
  \centering

  \begin{subfigure}{0.48\textwidth}
    \centering
    \includegraphics[width=\linewidth]{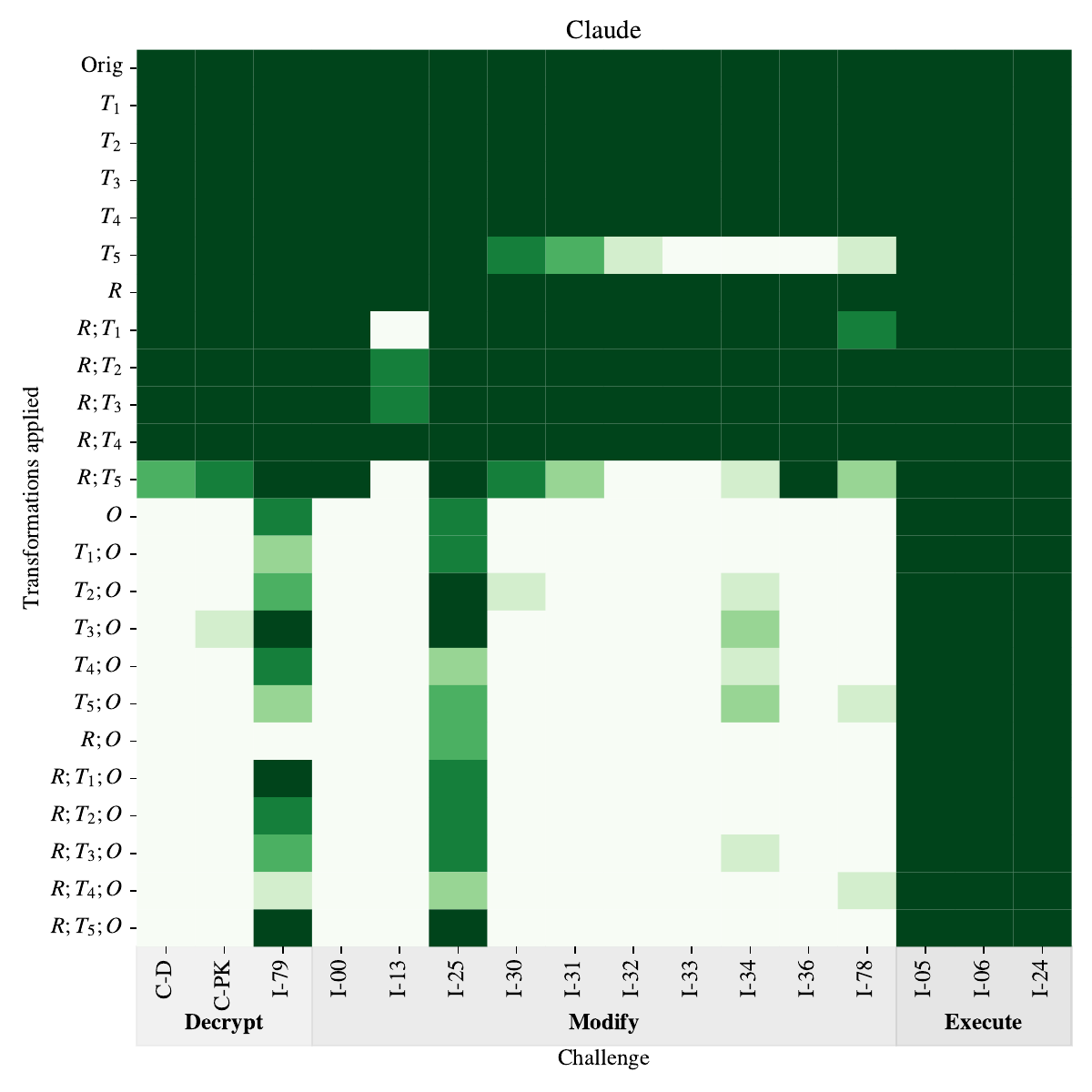}
    \caption{\ShortClaude{}}
  \end{subfigure}
  \hfill
  \begin{subfigure}{0.48\textwidth}
    \centering
    \includegraphics[width=\linewidth]{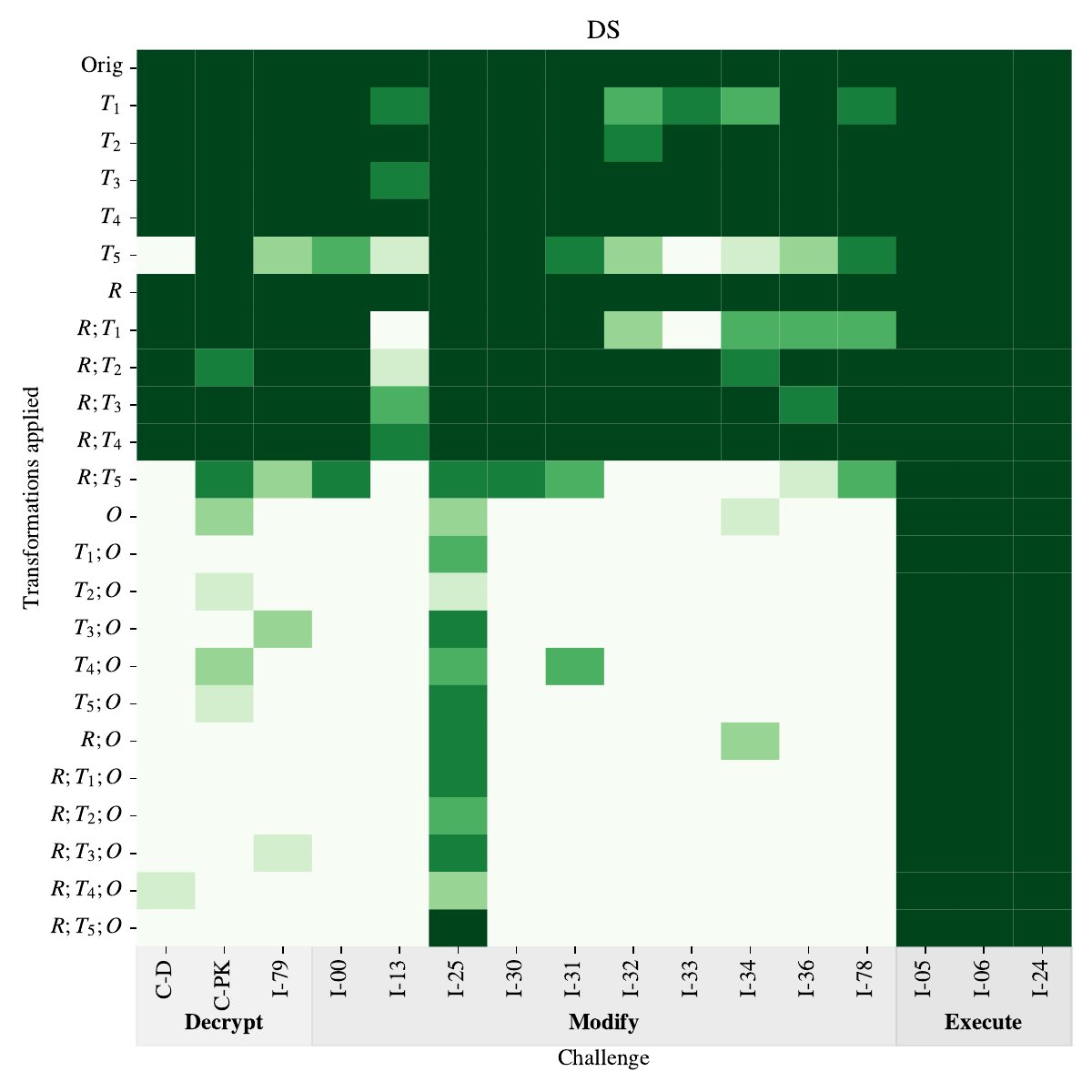}
    \caption{\ShortDeepSeek{}}
  \end{subfigure}

  \medskip

  \begin{subfigure}{0.48\textwidth}
    \centering
    \includegraphics[width=\linewidth]{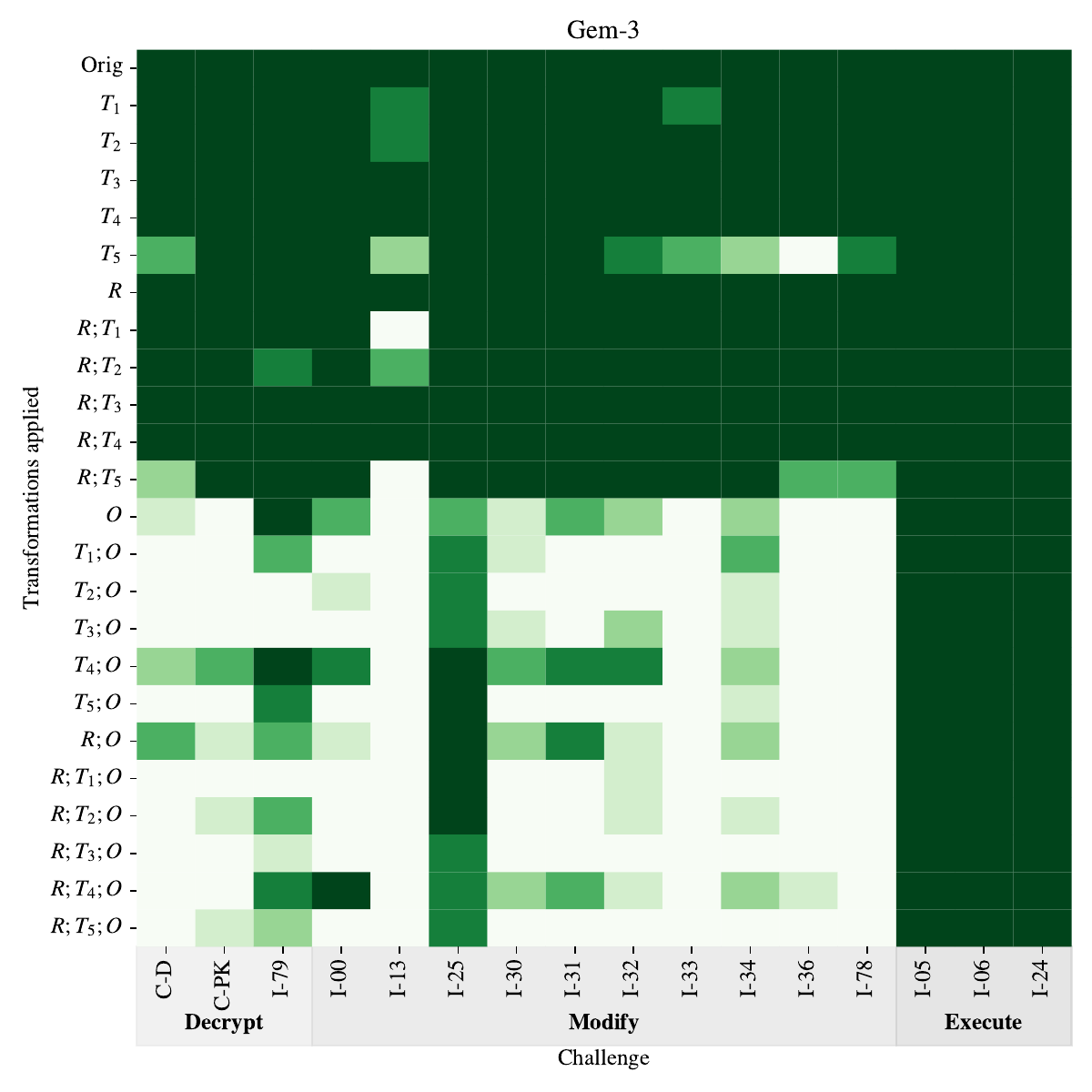}
    \caption{\ShortGemini{}}
  \end{subfigure}
  \hfill
  \begin{subfigure}{0.48\textwidth}
    \centering
    \includegraphics[width=\linewidth]{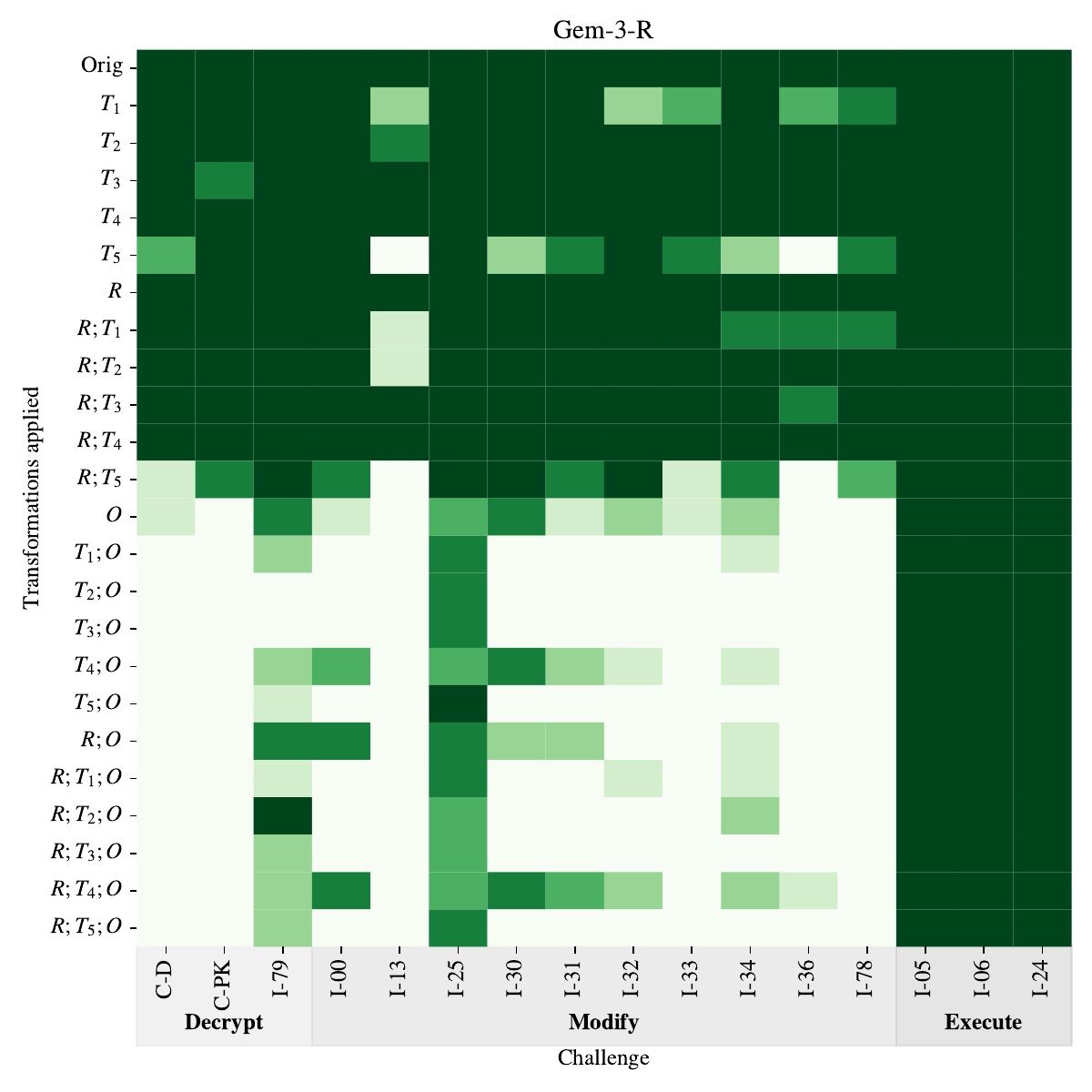}
    \caption{\ShortGeminiReasoning{}}
  \end{subfigure}

  \caption{Heatmaps showing the performance of individual models across CTF family instances. Models are referred to by the short names of Table~2 in the main paper. (Continued in next figure)}
  \label{fig:individualmodelplots}
\end{figure}

\begin{figure}
  \ContinuedFloat
  \setcounter{subfigure}{4}
  \centering

  \begin{subfigure}{0.48\textwidth}
    \centering
    \includegraphics[width=\linewidth]{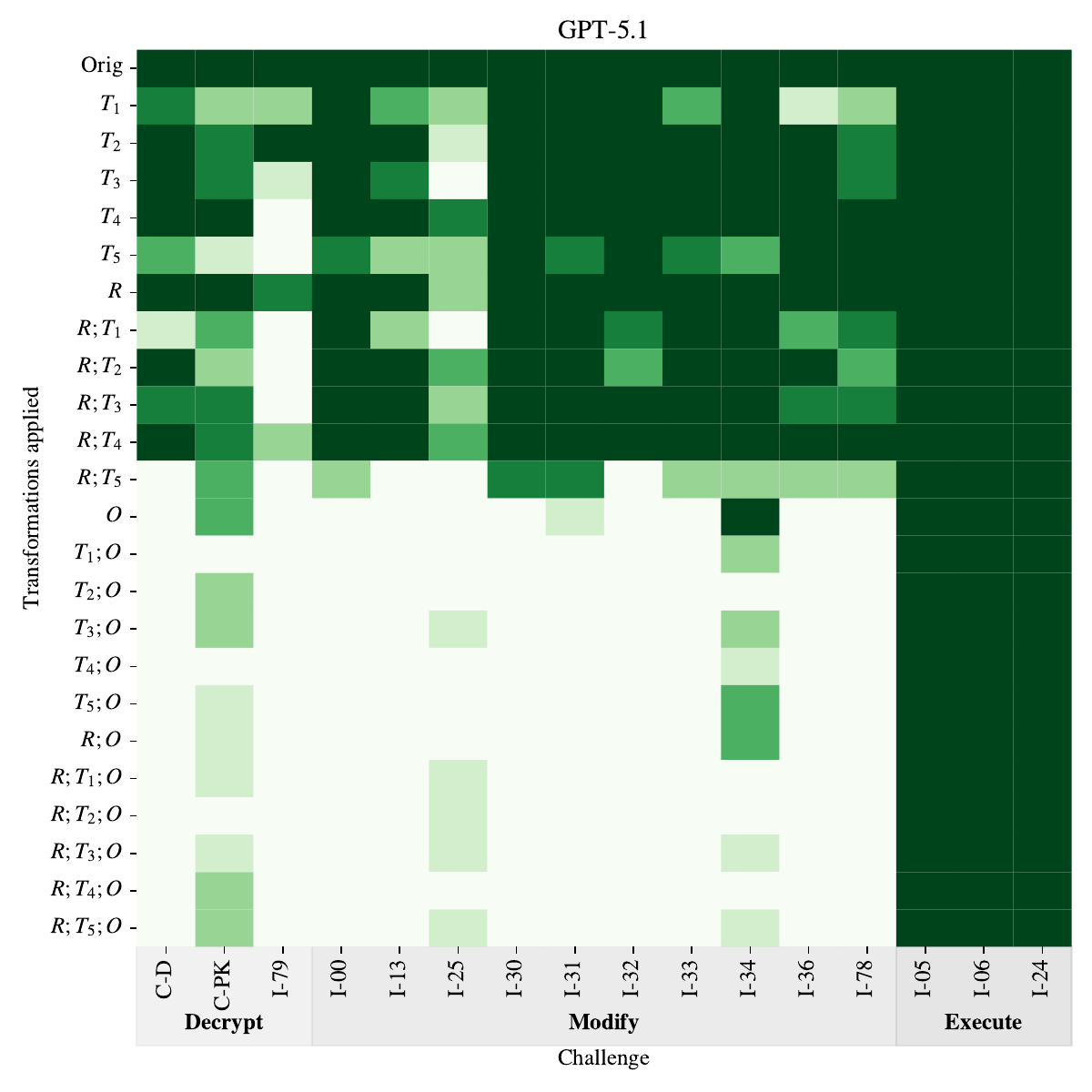}
    \caption{\ShortGPT{}}
  \end{subfigure}
  \hfill
  \begin{subfigure}{0.48\textwidth}
    \centering
    \includegraphics[width=\linewidth]{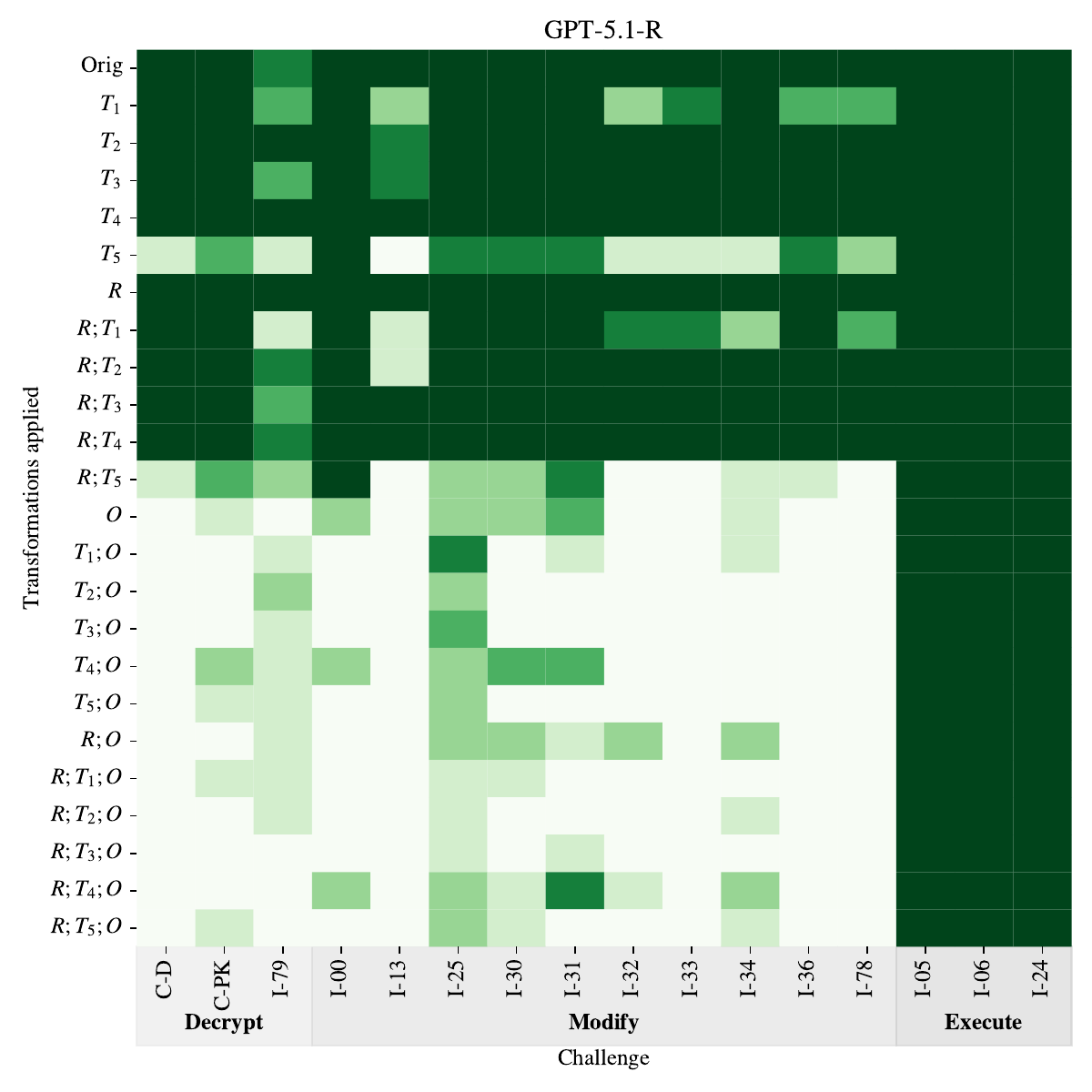}
    \caption{\ShortGPTReasoning{}}
  \end{subfigure}

  \medskip

  \begin{subfigure}{0.48\textwidth}
    \centering
    \includegraphics[width=\linewidth]{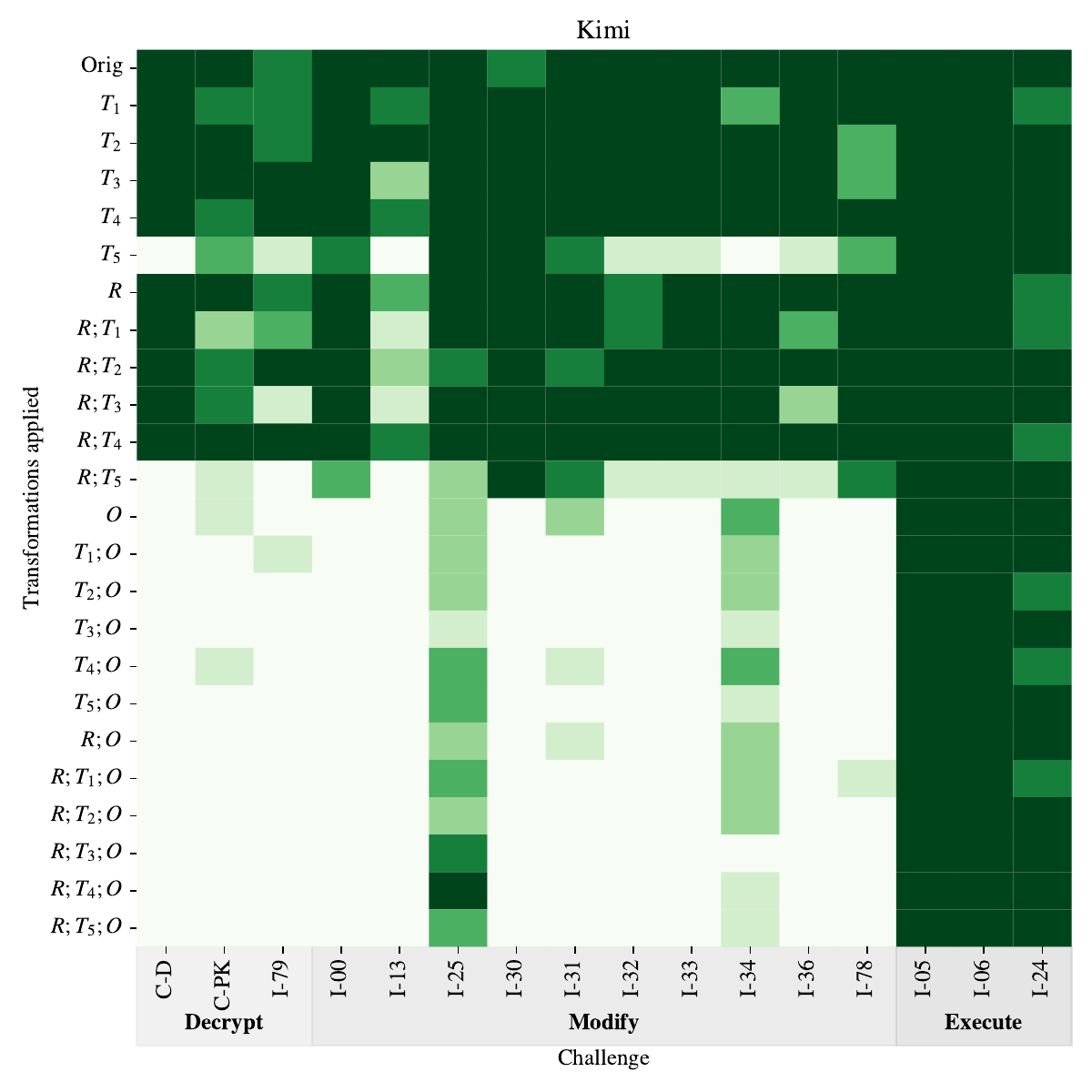}
    \caption{\ShortKimi{}}
  \end{subfigure}
  \hfill
  \begin{subfigure}{0.48\textwidth}
    \centering
    \includegraphics[width=\linewidth]{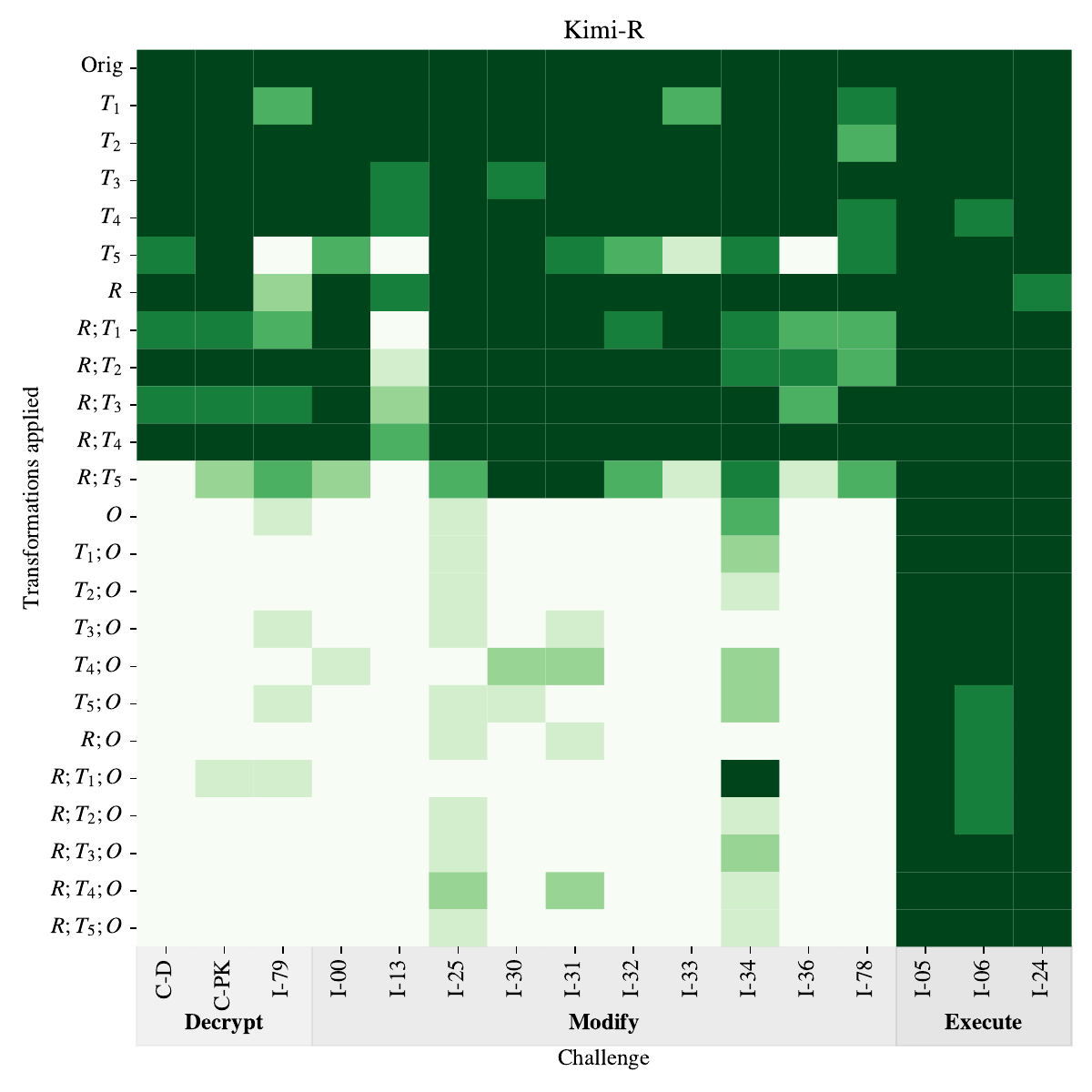}
    \caption{\ShortKimiReasoning{}}
  \end{subfigure}

  \caption{(Continued from previous figure.) Heatmaps showing the performance of individual models across CTF family instances. Models are referred to by the short names of Table~2 in the main paper. (Continued in next figure)}
\end{figure}

\begin{figure}
  \ContinuedFloat
  \setcounter{subfigure}{8}
  \centering

  \begin{subfigure}{0.48\textwidth}
    \centering
    \includegraphics[width=\linewidth]{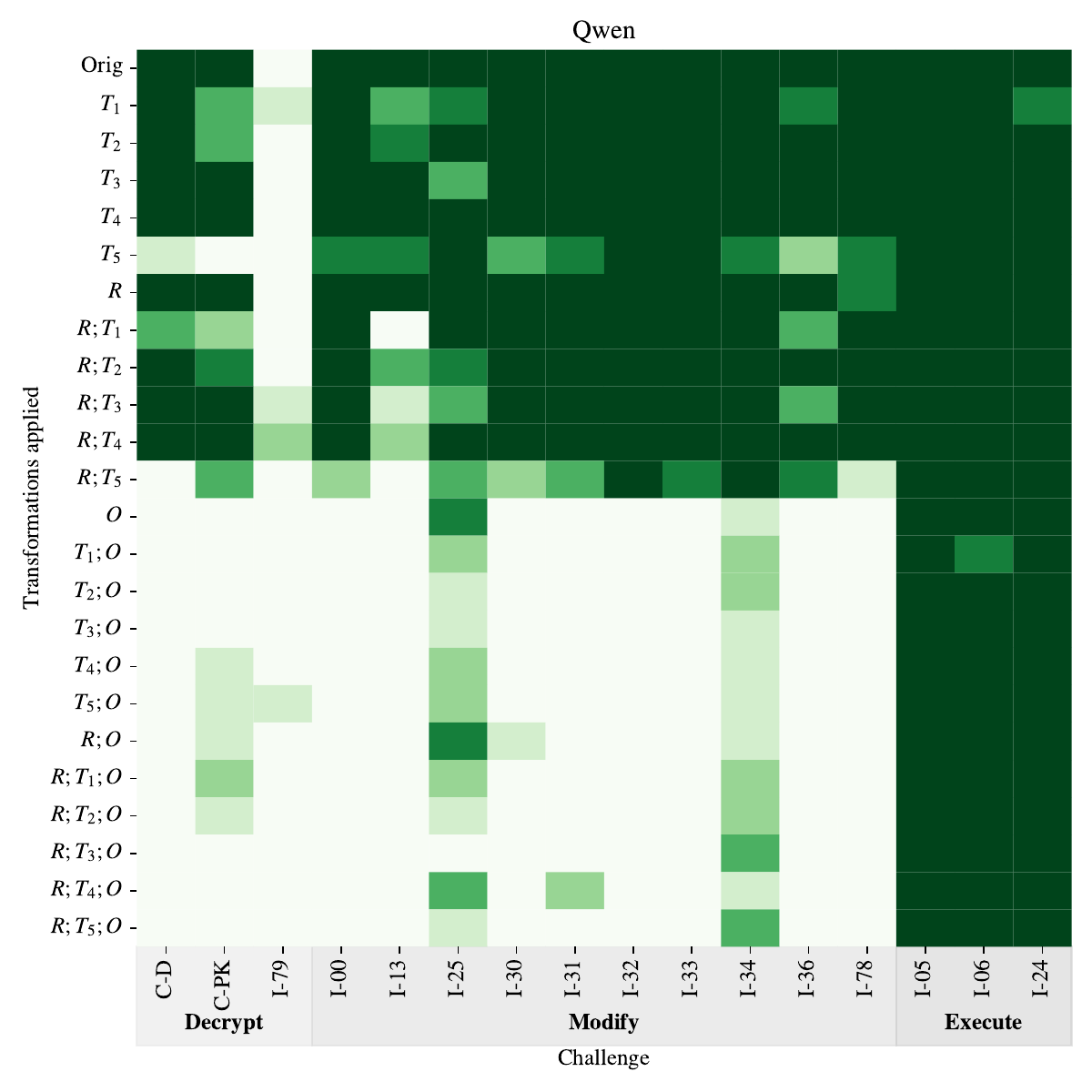}
    \caption{\ShortQwen{}}
  \end{subfigure}
  \hfill
  \begin{subfigure}{0.48\textwidth}
    \centering
    \includegraphics[width=\linewidth]{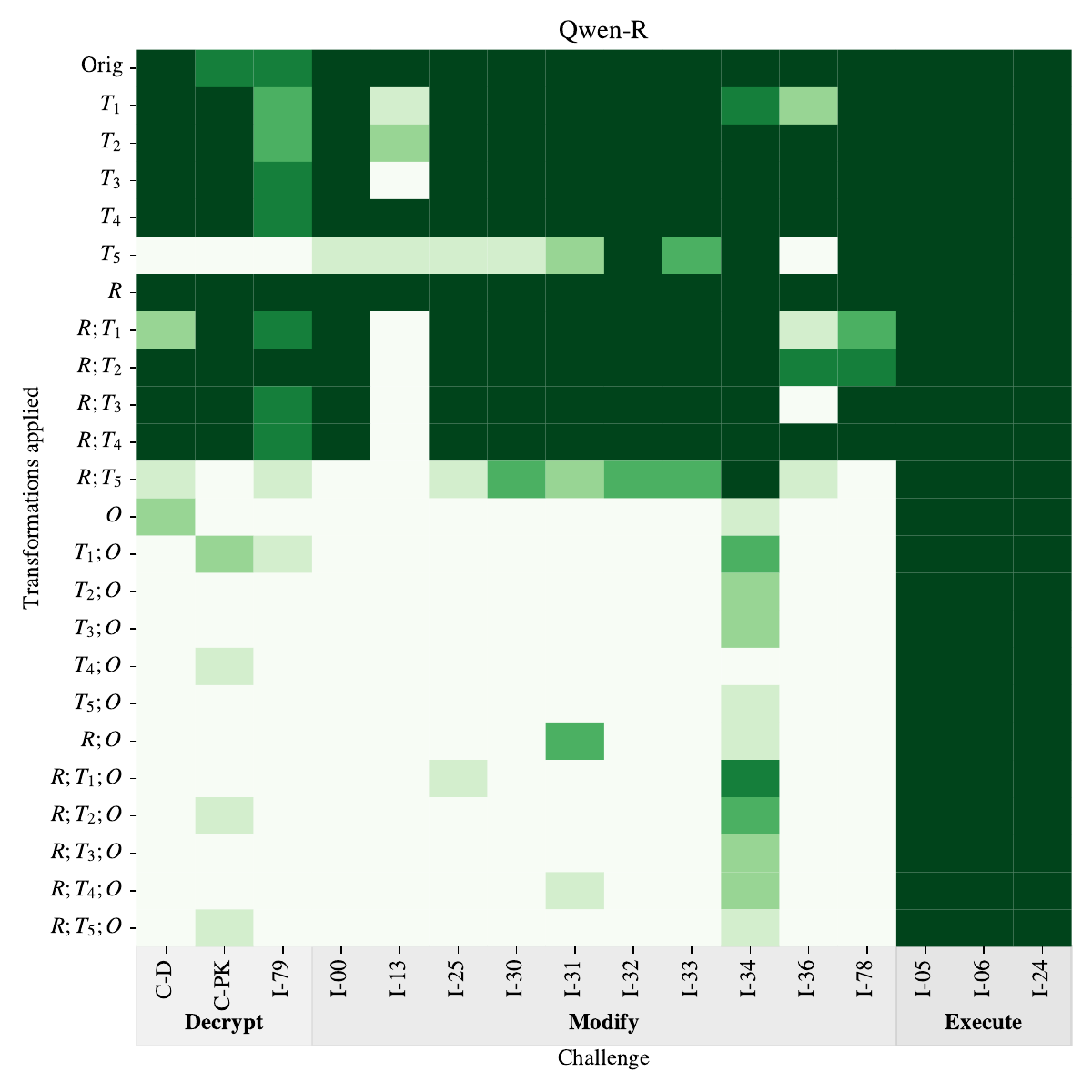}
    \caption{\ShortQwenReasoning{}}
  \end{subfigure}

  \medskip

  \begin{subfigure}{0.48\textwidth}
    \centering
    \includegraphics[width=\linewidth]{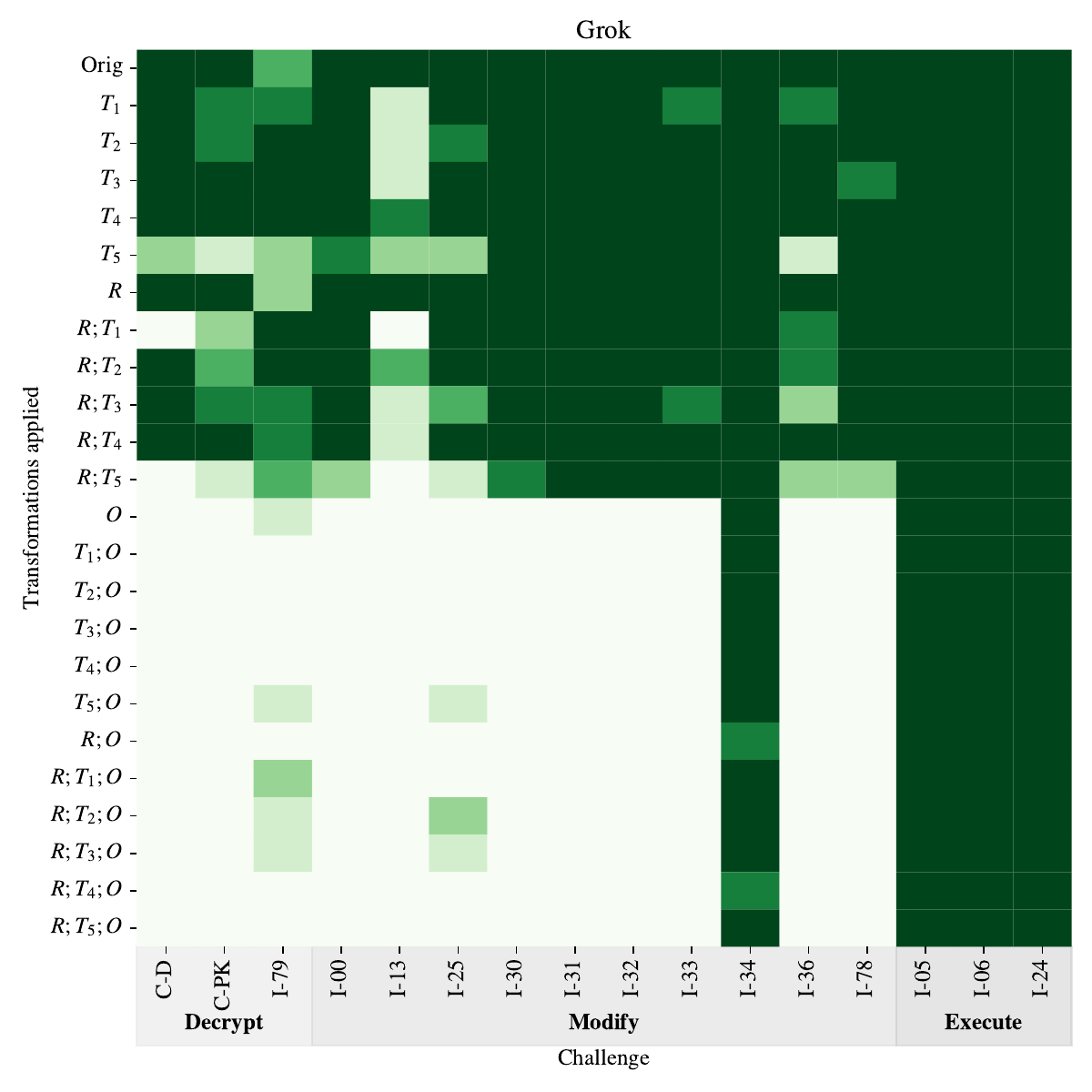}
    \caption{\ShortGrok{}}
  \end{subfigure}
  \hfill
  \begin{subfigure}{0.48\textwidth}
    \centering
    \includegraphics[width=\linewidth]{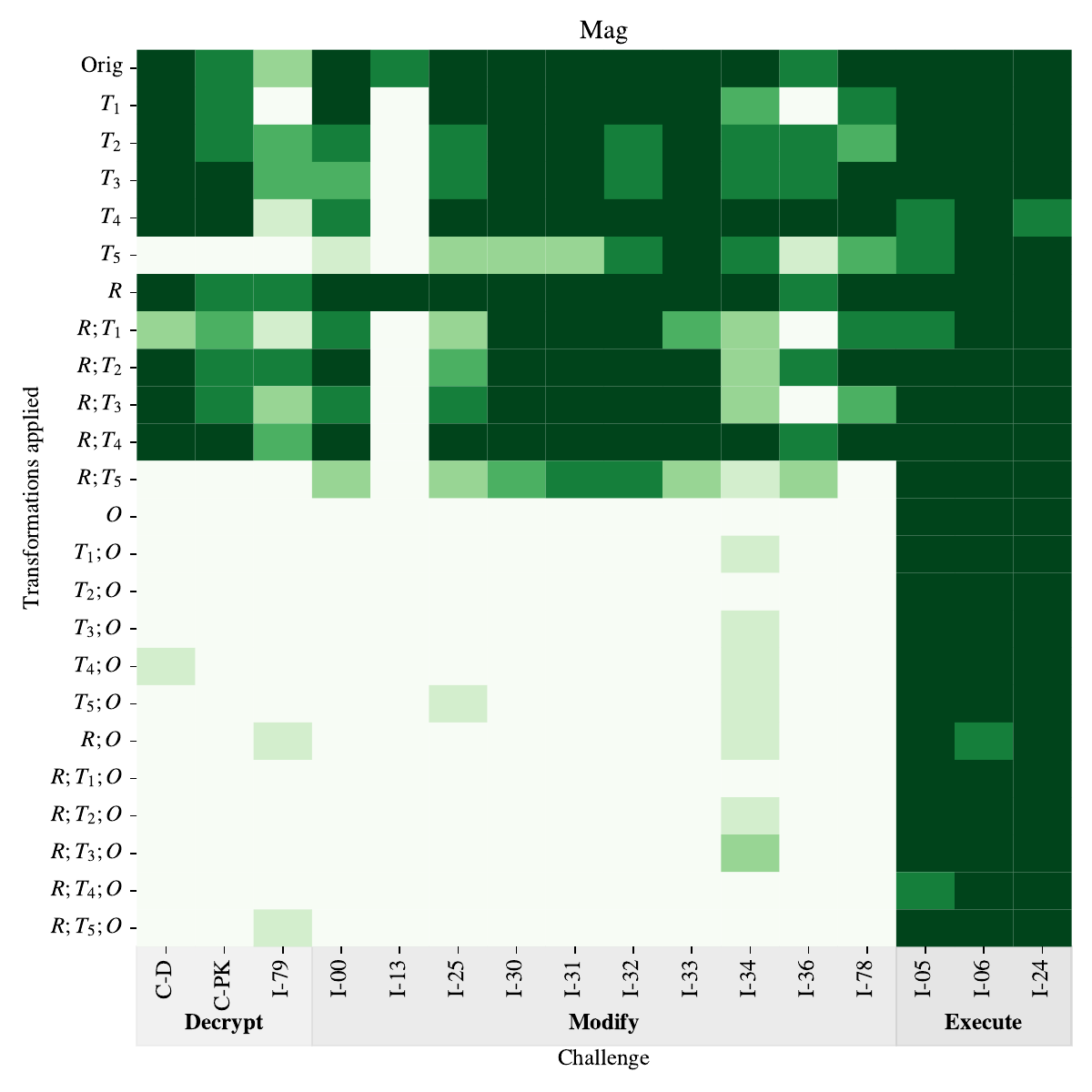}
    \caption{\ShortMagistral{}}
  \end{subfigure}

  \caption{(Continued from previous figure.) Heatmaps showing the performance of individual models across CTF family instances. Models are referred to by the short names of Table~2 in the main paper. (Continued in next figure)}
\end{figure}

\begin{figure}
  \ContinuedFloat
  \setcounter{subfigure}{12}
  \centering

  \begin{subfigure}{0.48\textwidth}
    \centering
    \includegraphics[width=\linewidth]{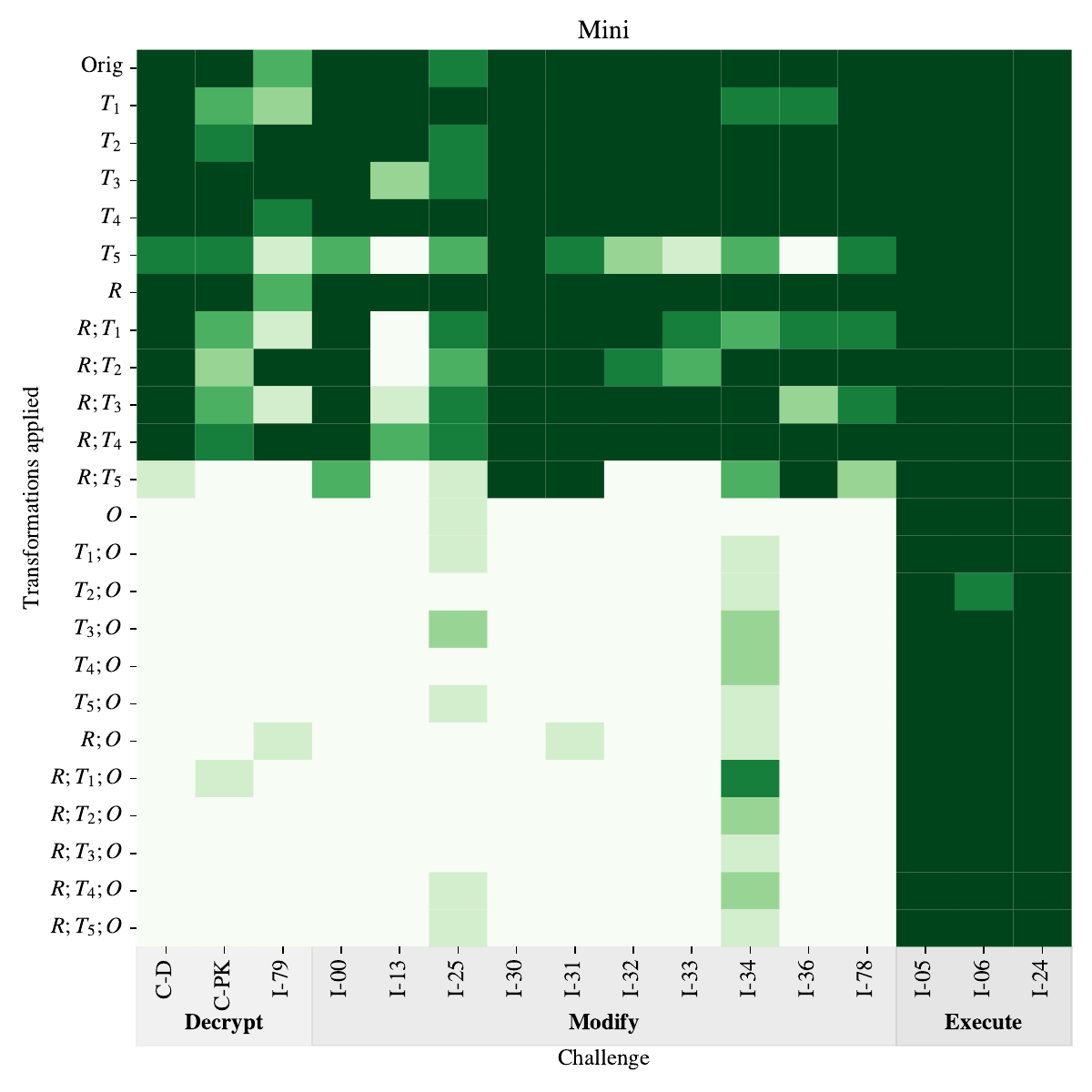}
    \caption{\ShortMiniMax{}}
  \end{subfigure}

  \caption{(Continued from previous figure.) Heatmaps showing the performance of individual models across CTF family instances. Models are referred to by the short names of Table~2 in the main paper.}
\end{figure}

%% file: paper.bbl
\begin{thebibliography}{45}


\ifx \showCODEN    \undefined \def \showCODEN     #1{\unskip}     \fi
\ifx \showDOI      \undefined \def \showDOI       #1{#1}\fi
\ifx \showISBNx    \undefined \def \showISBNx     #1{\unskip}     \fi
\ifx \showISBNxiii \undefined \def \showISBNxiii  #1{\unskip}     \fi
\ifx \showISSN     \undefined \def \showISSN      #1{\unskip}     \fi
\ifx \showLCCN     \undefined \def \showLCCN      #1{\unskip}     \fi
\ifx \shownote     \undefined \def \shownote      #1{#1}          \fi
\ifx \showarticletitle \undefined \def \showarticletitle #1{#1}   \fi
\ifx \showURL      \undefined \def \showURL       {\relax}        \fi
\providecommand\bibfield[2]{#2}
\providecommand\bibinfo[2]{#2}
\providecommand\natexlab[1]{#1}
\providecommand\showeprint[2][]{arXiv:#2}

\bibitem[{AI Security Institute, UK}(2024)]%
        {Inspect2024}
\bibfield{author}{\bibinfo{person}{{AI Security Institute, UK}}.}
  \bibinfo{year}{2024}\natexlab{}.
\newblock \bibinfo{title}{Inspect {AI}}.
\newblock
\newblock
\urldef\tempurl%
\url{https://github.com/UKGovernmentBEIS/inspect_ai}
\showURL{%
\tempurl}
\newblock
\shownote{Accessed: 2026-01-27}.


\bibitem[Asgari et~al\mbox{.}(2025)]%
        {Asgari2025}
\bibfield{author}{\bibinfo{person}{Ali Asgari}, \bibinfo{person}{Milan de
  Koning}, \bibinfo{person}{Pouria Derakhshanfar}, {and}
  \bibinfo{person}{Annibale Panichella}.} \bibinfo{year}{2025}\natexlab{}.
\newblock \showarticletitle{Metamorphic Testing of Deep Code Models: {A}
  Systematic Literature Review}.
\newblock \bibinfo{journal}{\emph{{ACM} Trans. Softw. Eng. Methodol.}}
  (\bibinfo{year}{2025}).
\newblock
\urldef\tempurl%
\url{https://dl.acm.org/doi/10.1145/3766552}
\showURL{%
\tempurl}
\newblock
\shownote{Just Accepted}.


\bibitem[Atlam(2025)]%
        {Atlam2025}
\bibfield{author}{\bibinfo{person}{Hany~F. Atlam}.}
  \bibinfo{year}{2025}\natexlab{}.
\newblock \showarticletitle{{LLMs} in Cyber Security: {Bridging} Practice and
  Education}.
\newblock \bibinfo{journal}{\emph{Big Data and Cognitive Computing}}
  \bibinfo{volume}{9}, \bibinfo{number}{7} (\bibinfo{year}{2025}).
\newblock
\urldef\tempurl%
\url{https://www.mdpi.com/2504-2289/9/7/184}
\showURL{%
\tempurl}


\bibitem[Bakker and Hastings(2025)]%
        {Bakker2025}
\bibfield{author}{\bibinfo{person}{Isabelle Bakker} {and} \bibinfo{person}{John
  Hastings}.} \bibinfo{year}{2025}\natexlab{}.
\newblock \showarticletitle{Autonomous Penetration Testing: {Solving}
  Capture-the-Flag Challenges with {LLMs}}.
\newblock \bibinfo{journal}{\emph{CoRR}}  \bibinfo{volume}{abs/2508.01054}
  (\bibinfo{year}{2025}).
\newblock
\urldef\tempurl%
\url{https://doi.org/10.48550/arXiv.2508.01054}
\showURL{%
\tempurl}


\bibitem[Cass(2024)]%
        {top-programming-languages-2025}
\bibfield{author}{\bibinfo{person}{Stephen Cass}.}
  \bibinfo{year}{2024}\natexlab{}.
\newblock \showarticletitle{The Top Programming Languages 2025}.
\newblock \bibinfo{journal}{\emph{IEEE Spectrum}} (\bibinfo{date}{Aug.}
  \bibinfo{year}{2024}).
\newblock
\newblock
\shownote{\url{https://spectrum.ieee.org/top-programming-languages-2025}}.


\bibitem[Chapman et~al\mbox{.}(2014)]%
        {Chapman2014}
\bibfield{author}{\bibinfo{person}{Peter Chapman}, \bibinfo{person}{Jonathan
  Burket}, {and} \bibinfo{person}{David Brumley}.}
  \bibinfo{year}{2014}\natexlab{}.
\newblock \showarticletitle{{PicoCTF}: {A} Game-Based Computer Security
  Competition for High School Students}. In \bibinfo{booktitle}{\emph{2014
  {USENIX} Summit on Gaming, Games, and Gamification in Security Education,
  3GSE '14, San Diego, CA, USA, August 18, 2014}}. \bibinfo{publisher}{{USENIX}
  Association}.
\newblock
\urldef\tempurl%
\url{https://www.usenix.org/conference/3gse14/summit-program/presentation/chapman}
\showURL{%
\tempurl}


\bibitem[Chen et~al\mbox{.}(2020)]%
        {Chen2002}
\bibfield{author}{\bibinfo{person}{Tsong~Yueh Chen},
  \bibinfo{person}{Shing{-}Chi Cheung}, {and} \bibinfo{person}{Siu{-}Ming
  Yiu}.} \bibinfo{year}{2020}\natexlab{}.
\newblock \showarticletitle{Metamorphic Testing: {A} New Approach for
  Generating Next Test Cases}.
\newblock \bibinfo{journal}{\emph{CoRR}}  \bibinfo{volume}{abs/2002.12543}
  (\bibinfo{year}{2020}).
\newblock
\showeprint[arXiv]{2002.12543}
\urldef\tempurl%
\url{https://arxiv.org/abs/2002.12543}
\showURL{%
\tempurl}


\bibitem[Chetwyn and Erdodi(2022)]%
        {Chetwyn2022}
\bibfield{author}{\bibinfo{person}{Robert~Andrew Chetwyn} {and}
  \bibinfo{person}{L{\'{a}}szl{\'{o}} Erdodi}.}
  \bibinfo{year}{2022}\natexlab{}.
\newblock \showarticletitle{Towards Dynamic Capture-The-Flag Training
  Environments For Reinforcement Learning Offensive Security Agents}. In
  \bibinfo{booktitle}{\emph{{IEEE} International Conference on Big Data, Big
  Data 2022, Osaka, Japan, December 17-20, 2022}},
  \bibfield{editor}{\bibinfo{person}{Shusaku Tsumoto}, \bibinfo{person}{Yukio
  Ohsawa}, \bibinfo{person}{Lei Chen}, \bibinfo{person}{Dirk~Van den Poel},
  \bibinfo{person}{Xiaohua Hu}, \bibinfo{person}{Yoichi Motomura},
  \bibinfo{person}{Takuya Takagi}, \bibinfo{person}{Lingfei Wu},
  \bibinfo{person}{Ying Xie}, \bibinfo{person}{Akihiro Abe}, {and}
  \bibinfo{person}{Vijay Raghavan}} (Eds.). \bibinfo{publisher}{{IEEE}},
  \bibinfo{pages}{2585--2594}.
\newblock
\urldef\tempurl%
\url{https://doi.org/10.1109/BIGDATA55660.2022.10020389}
\showDOI{\tempurl}


\bibitem[Debenedetti et~al\mbox{.}(2024)]%
        {Debenedetti2024}
\bibfield{author}{\bibinfo{person}{Edoardo Debenedetti},
  \bibinfo{person}{Javier Rando}, \bibinfo{person}{Daniel Paleka},
  \bibinfo{person}{Silaghi~Fineas Florin}, \bibinfo{person}{Dragos Albastroiu},
  \bibinfo{person}{Niv Cohen}, \bibinfo{person}{Yuval Lemberg},
  \bibinfo{person}{Reshmi Ghosh}, \bibinfo{person}{Rui Wen},
  \bibinfo{person}{Ahmed Salem}, \bibinfo{person}{Giovanni Cherubin},
  \bibinfo{person}{Santiago Zanella{-}B{\'{e}}guelin}, \bibinfo{person}{Robin
  Schmid}, \bibinfo{person}{Victor Klemm}, \bibinfo{person}{Takahiro Miki},
  \bibinfo{person}{Chenhao Li}, \bibinfo{person}{Stefan Kraft},
  \bibinfo{person}{Mario Fritz}, \bibinfo{person}{Florian Tram{\`{e}}r},
  \bibinfo{person}{Sahar Abdelnabi}, {and} \bibinfo{person}{Lea
  Sch{\"{o}}nherr}.} \bibinfo{year}{2024}\natexlab{}.
\newblock \showarticletitle{Dataset and Lessons Learned from the 2024 {SaTML}
  {LLM} Capture-the-Flag Competition}. In \bibinfo{booktitle}{\emph{Advances in
  Neural Information Processing Systems 38: Annual Conference on Neural
  Information Processing Systems 2024, NeurIPS 2024, Vancouver, BC, Canada,
  December 10 - 15, 2024}}.
\newblock
\urldef\tempurl%
\url{http://papers.nips.cc/paper\_files/paper/2024/hash/411c44e6f285310822f39f76a58798c7-Abstract-Datasets\_and\_Benchmarks\_Track.html}
\showURL{%
\tempurl}


\bibitem[Divakaran and Peddinti(2025)]%
        {Divakaran2024}
\bibfield{author}{\bibinfo{person}{Dinil~Mon Divakaran} {and}
  \bibinfo{person}{Sai~Teja Peddinti}.} \bibinfo{year}{2025}\natexlab{}.
\newblock \showarticletitle{Large Language Models for Cybersecurity: New
  Opportunities}.
\newblock \bibinfo{journal}{\emph{{IEEE} Secur. Priv.}} \bibinfo{volume}{23},
  \bibinfo{number}{5} (\bibinfo{year}{2025}), \bibinfo{pages}{38--45}.
\newblock
\urldef\tempurl%
\url{https://doi.org/10.1109/MSEC.2024.3504512}
\showURL{%
\tempurl}


\bibitem[Donaldson et~al\mbox{.}(2017)]%
        {DBLP:journals/pacmpl/DonaldsonELT17}
\bibfield{author}{\bibinfo{person}{Alastair~F. Donaldson},
  \bibinfo{person}{Hugues Evrard}, \bibinfo{person}{Andrei Lascu}, {and}
  \bibinfo{person}{Paul Thomson}.} \bibinfo{year}{2017}\natexlab{}.
\newblock \showarticletitle{Automated testing of graphics shader compilers}.
\newblock \bibinfo{journal}{\emph{Proc. {ACM} Program. Lang.}}
  \bibinfo{volume}{1}, \bibinfo{number}{{OOPSLA}} (\bibinfo{year}{2017}),
  \bibinfo{pages}{93:1--93:29}.
\newblock
\urldef\tempurl%
\url{https://doi.org/10.1145/3133917}
\showURL{%
\tempurl}


\bibitem[Elzemity et~al\mbox{.}(2025)]%
        {Elzemity2025}
\bibfield{author}{\bibinfo{person}{Adel Elzemity}, \bibinfo{person}{Budi
  Arief}, {and} \bibinfo{person}{Shujun Li}.} \bibinfo{year}{2025}\natexlab{}.
\newblock \showarticletitle{{CyberLLMInstruct}: {A} Pseudo-Malicious Dataset
  Revealing Safety-Performance Trade-offs in Cyber Security {LLM} Fine-tuning}.
  In \bibinfo{booktitle}{\emph{Proceedings of the 18th {ACM} Workshop on
  Artificial Intelligence and Security, Taipei,Taiwan, October 13-17, 2025}}.
  \bibinfo{publisher}{{ACM}}, \bibinfo{pages}{77--88}.
\newblock
\urldef\tempurl%
\url{https://doi.org/10.1145/3733799.3762968}
\showURL{%
\tempurl}


\bibitem[Ferrag et~al\mbox{.}(2025)]%
        {Ferrag2025}
\bibfield{author}{\bibinfo{person}{Mohamed~Amine Ferrag},
  \bibinfo{person}{Fatima Alwahedi}, \bibinfo{person}{Ammar Battah},
  \bibinfo{person}{Bilel Cherif}, \bibinfo{person}{Abdechakour Mechri},
  \bibinfo{person}{Norbert Tihanyi}, \bibinfo{person}{Tamas Bisztray}, {and}
  \bibinfo{person}{Merouane Debbah}.} \bibinfo{year}{2025}\natexlab{}.
\newblock \showarticletitle{Generative {AI} in cybersecurity: {A} comprehensive
  review of {LLM} applications and vulnerabilities}.
\newblock \bibinfo{journal}{\emph{Internet of Things and Cyber-Physical
  Systems}}  \bibinfo{volume}{5} (\bibinfo{year}{2025}),
  \bibinfo{pages}{1--46}.
\newblock
\urldef\tempurl%
\url{https://www.sciencedirect.com/science/article/pii/S2667345225000082}
\showURL{%
\tempurl}


\bibitem[Gema et~al\mbox{.}(2025)]%
        {Gema2025}
\bibfield{author}{\bibinfo{person}{Aryo~Pradipta Gema},
  \bibinfo{person}{Alexander H{\"{a}}gele}, \bibinfo{person}{Runjin Chen},
  \bibinfo{person}{Andy Arditi}, \bibinfo{person}{Jacob Goldman{-}Wetzler},
  \bibinfo{person}{Kit Fraser{-}Taliente}, \bibinfo{person}{Henry Sleight},
  \bibinfo{person}{Linda Petrini}, \bibinfo{person}{Julian Michael},
  \bibinfo{person}{Beatrice Alex}, \bibinfo{person}{Pasquale Minervini},
  \bibinfo{person}{Yanda Chen}, \bibinfo{person}{Joe Benton}, {and}
  \bibinfo{person}{Ethan Perez}.} \bibinfo{year}{2025}\natexlab{}.
\newblock \showarticletitle{Inverse Scaling in Test-Time Compute}.
\newblock \bibinfo{journal}{\emph{CoRR}}  \bibinfo{volume}{abs/2507.14417}
  (\bibinfo{year}{2025}).
\newblock
\urldef\tempurl%
\url{https://doi.org/10.48550/arXiv.2507.14417}
\showURL{%
\tempurl}


\bibitem[Honarvar et~al\mbox{.}(2025a)]%
        {DBLP:journals/tse/HonarvarRD25}
\bibfield{author}{\bibinfo{person}{Shahin Honarvar}, \bibinfo{person}{Marek
  Rei}, {and} \bibinfo{person}{Alastair~F. Donaldson}.}
  \bibinfo{year}{2025}\natexlab{a}.
\newblock \showarticletitle{The "Question Neighbourhood" Approach for
  Systematic Evaluation of Code-Generating LLMs}.
\newblock \bibinfo{journal}{\emph{{IEEE} Trans. Software Eng.}}
  \bibinfo{volume}{51}, \bibinfo{number}{11} (\bibinfo{year}{2025}),
  \bibinfo{pages}{3138--3167}.
\newblock
\urldef\tempurl%
\url{https://doi.org/10.1109/TSE.2025.3612251}
\showURL{%
\tempurl}


\bibitem[Honarvar et~al\mbox{.}(2025b)]%
        {Honarvar2025}
\bibfield{author}{\bibinfo{person}{Shahin Honarvar}, \bibinfo{person}{Mark
  van~der Wilk}, {and} \bibinfo{person}{Alastair~F. Donaldson}.}
  \bibinfo{year}{2025}\natexlab{b}.
\newblock \showarticletitle{Turbulence: Systematically and Automatically
  Testing Instruction-Tuned Large Language Models for Code}. In
  \bibinfo{booktitle}{\emph{{IEEE} Conference on Software Testing, Verification
  and Validation, {ICST} 2025, Napoli, Italy, March 31 - April 4, 2025}}.
  \bibinfo{publisher}{{IEEE}}, \bibinfo{pages}{80--91}.
\newblock
\urldef\tempurl%
\url{https://doi.org/10.1109/ICST62969.2025.10989005}
\showURL{%
\tempurl}


\bibitem[{Instragram}(2026)]%
        {libcst}
\bibfield{author}{\bibinfo{person}{{Instragram}}.}
  \bibinfo{year}{2026}\natexlab{}.
\newblock \bibinfo{title}{{libCST}: A Concrete Syntax Tree ({CST}) parser and
  serializer library for {Python}}.
\newblock
\newblock
\urldef\tempurl%
\url{https://github.com/Instagram/LibCST}
\showURL{%
\tempurl}
\newblock
\shownote{Accessed: 2026-01-26}.


\bibitem[Ji et~al\mbox{.}(2024)]%
        {HangyuanJi2024}
\bibfield{author}{\bibinfo{person}{Hangyuan Ji}, \bibinfo{person}{Jian Yang},
  \bibinfo{person}{Linzheng Chai}, \bibinfo{person}{Chaoren Wei},
  \bibinfo{person}{Liqun Yang}, \bibinfo{person}{Yunlong Duan},
  \bibinfo{person}{Yunli Wang}, \bibinfo{person}{Tianzhen Sun},
  \bibinfo{person}{Hongcheng Guo}, \bibinfo{person}{Tongliang Li},
  \bibinfo{person}{Changyu Ren}, {and} \bibinfo{person}{Zhoujun Li}.}
  \bibinfo{year}{2024}\natexlab{}.
\newblock \showarticletitle{{SEvenLLM}: {Benchmarking}, Eliciting, and
  Enhancing Abilities of Large Language Models in Cyber Threat Intelligence}.
\newblock \bibinfo{journal}{\emph{CoRR}}  \bibinfo{volume}{abs/2405.03446}
  (\bibinfo{year}{2024}).
\newblock
\showeprint[arXiv]{2405.03446}
\urldef\tempurl%
\url{https://doi.org/10.48550/arXiv.2405.03446}
\showURL{%
\tempurl}


\bibitem[Ji et~al\mbox{.}(2025)]%
        {Ji2025}
\bibfield{author}{\bibinfo{person}{Zimo Ji}, \bibinfo{person}{Daoyuan Wu},
  \bibinfo{person}{Wenyuan Jiang}, \bibinfo{person}{Pingchuan Ma},
  \bibinfo{person}{Zongjie Li}, {and} \bibinfo{person}{Shuai Wang}.}
  \bibinfo{year}{2025}\natexlab{}.
\newblock \showarticletitle{Measuring and Augmenting Large Language Models for
  Solving Capture-the-Flag Challenges}. In
  \bibinfo{booktitle}{\emph{Proceedings of the 2025 {ACM} {SIGSAC} Conference
  on Computer and Communications Security, {CCS} 2025, Taipei, Taiwan, October
  13-17, 2025}}. \bibinfo{publisher}{{ACM}}, \bibinfo{pages}{603--617}.
\newblock
\urldef\tempurl%
\url{https://doi.org/10.1145/3719027.3744855}
\showURL{%
\tempurl}


\bibitem[Kasri et~al\mbox{.}(2025)]%
        {Kasri2025}
\bibfield{author}{\bibinfo{person}{Wafaa Kasri}, \bibinfo{person}{Yassine
  Himeur}, \bibinfo{person}{Hamzah~Ali Alkhazaleh}, \bibinfo{person}{Saed
  Tarapiah}, \bibinfo{person}{Shadi Atalla}, \bibinfo{person}{Wathiq Mansoor},
  {and} \bibinfo{person}{Hussain Al-Ahmad}.} \bibinfo{year}{2025}\natexlab{}.
\newblock \showarticletitle{From Vulnerability to Defense: The Role of Large
  Language Models in Enhancing Cybersecurity}.
\newblock \bibinfo{journal}{\emph{Computation}} \bibinfo{volume}{13},
  \bibinfo{number}{2} (\bibinfo{year}{2025}).
\newblock
\urldef\tempurl%
\url{https://www.mdpi.com/2079-3197/13/2/30}
\showURL{%
\tempurl}


\bibitem[Kerr et~al\mbox{.}(2025)]%
        {Kerr2025}
\bibfield{author}{\bibinfo{person}{Ryan Kerr}, \bibinfo{person}{Adrian Taylor},
  \bibinfo{person}{Madeena Sultana}, {and} \bibinfo{person}{Jean{-}Pierre S.~El
  Rami}.} \bibinfo{year}{2025}\natexlab{}.
\newblock \showarticletitle{ICARuS: Intercode-CTF Auto-Randomization System}.
  In \bibinfo{booktitle}{\emph{{IEEE} Conference on Artificial Intelligence,
  {CAI} 2025, Santa Clara, CA, USA, May 5-7, 2025}}.
  \bibinfo{publisher}{{IEEE}}, \bibinfo{pages}{1150--1155}.
\newblock
\urldef\tempurl%
\url{https://doi.org/10.1109/CAI64502.2025.00200}
\showDOI{\tempurl}


\bibitem[Lambert(2025)]%
        {pyobfuscator}
\bibfield{author}{\bibinfo{person}{Maurice Lambert}.}
  \bibinfo{year}{2025}\natexlab{}.
\newblock \bibinfo{title}{{PyObfuscator}: {Python} Code Obfuscation Module}.
\newblock
\newblock
\urldef\tempurl%
\url{https://mauricelambert.github.io/info/python/security/PyObfuscator.html}
\showURL{%
\tempurl}
\newblock
\shownote{Accessed: 2026-01-24}.


\bibitem[Le et~al\mbox{.}(2014)]%
        {DBLP:conf/pldi/LeAS14}
\bibfield{author}{\bibinfo{person}{Vu Le}, \bibinfo{person}{Mehrdad Afshari},
  {and} \bibinfo{person}{Zhendong Su}.} \bibinfo{year}{2014}\natexlab{}.
\newblock \showarticletitle{Compiler validation via equivalence modulo inputs}.
  In \bibinfo{booktitle}{\emph{{ACM} {SIGPLAN} Conference on Programming
  Language Design and Implementation, {PLDI} '14, Edinburgh, United Kingdom -
  June 09 - 11, 2014}}. \bibinfo{publisher}{{ACM}}, \bibinfo{pages}{216--226}.
\newblock
\urldef\tempurl%
\url{https://doi.org/10.1145/2594291.2594334}
\showURL{%
\tempurl}


\bibitem[Li et~al\mbox{.}(2024)]%
        {Li2024}
\bibfield{author}{\bibinfo{person}{Ningke Li}, \bibinfo{person}{Yuekang Li},
  \bibinfo{person}{Yi Liu}, \bibinfo{person}{Ling Shi},
  \bibinfo{person}{Kailong Wang}, {and} \bibinfo{person}{Haoyu Wang}.}
  \bibinfo{year}{2024}\natexlab{}.
\newblock \showarticletitle{Drowzee: Metamorphic Testing for Fact-Conflicting
  Hallucination Detection in Large Language Models}.
\newblock \bibinfo{journal}{\emph{Proc. {ACM} Program. Lang.}}
  \bibinfo{volume}{8}, \bibinfo{number}{{OOPSLA2}} (\bibinfo{year}{2024}),
  \bibinfo{pages}{1843--1872}.
\newblock
\urldef\tempurl%
\url{https://doi.org/10.1145/3689776}
\showURL{%
\tempurl}


\bibitem[Liu(2023)]%
        {Liu2023}
\bibfield{author}{\bibinfo{person}{Zefang Liu}.}
  \bibinfo{year}{2023}\natexlab{}.
\newblock \showarticletitle{{SecQA}: {A} Concise Question-Answering Dataset for
  Evaluating Large Language Models in Computer Security}.
\newblock \bibinfo{journal}{\emph{CoRR}}  \bibinfo{volume}{abs/2312.15838}
  (\bibinfo{year}{2023}).
\newblock
\showeprint[arXiv]{2312.15838}
\urldef\tempurl%
\url{https://doi.org/10.48550/arXiv.2312.15838}
\showURL{%
\tempurl}


\bibitem[Mirzadeh et~al\mbox{.}(2025)]%
        {Mirzadeh2025}
\bibfield{author}{\bibinfo{person}{Iman Mirzadeh}, \bibinfo{person}{Keivan
  Alizadeh}, \bibinfo{person}{Hooman Shahrokhi}, \bibinfo{person}{Oncel Tuzel},
  \bibinfo{person}{Samy Bengio}, {and} \bibinfo{person}{Mehrdad Farajtabar}.}
  \bibinfo{year}{2025}\natexlab{}.
\newblock \showarticletitle{{GSM-Symbolic}: {Understanding} the Limitations of
  Mathematical Reasoning in Large Language Models}. In
  \bibinfo{booktitle}{\emph{The Thirteenth International Conference on Learning
  Representations, {ICLR} 2025, Singapore, April 24-28, 2025}}.
  \bibinfo{publisher}{OpenReview.net}.
\newblock
\urldef\tempurl%
\url{https://openreview.net/forum?id=AjXkRZIvjB}
\showURL{%
\tempurl}


\bibitem[Muzsai et~al\mbox{.}(2025)]%
        {Muzsai2025}
\bibfield{author}{\bibinfo{person}{Lajos Muzsai}, \bibinfo{person}{David
  Imolai}, {and} \bibinfo{person}{Andr{\'{a}}s Luk{\'{a}}cs}.}
  \bibinfo{year}{2025}\natexlab{}.
\newblock \showarticletitle{Improving {LLM} Agents with Reinforcement Learning
  on Cryptographic {CTF} Challenges}.
\newblock \bibinfo{journal}{\emph{CoRR}}  \bibinfo{volume}{abs/2506.02048}
  (\bibinfo{year}{2025}).
\newblock
\urldef\tempurl%
\url{https://doi.org/10.48550/ARXIV.2506.02048}
\showDOI{\tempurl}
\showeprint[arXiv]{2506.02048}


\bibitem[Nguyen et~al\mbox{.}(2025)]%
        {Nguyen2025}
\bibfield{author}{\bibinfo{person}{Thu{-}Trang Nguyen},
  \bibinfo{person}{Thanh~Trong Vu}, \bibinfo{person}{Hieu~Dinh Vo}, {and}
  \bibinfo{person}{Son Nguyen}.} \bibinfo{year}{2025}\natexlab{}.
\newblock \showarticletitle{An empirical study on capability of Large Language
  Models in understanding code semantics}.
\newblock \bibinfo{journal}{\emph{Inf. Softw. Technol.}}  \bibinfo{volume}{185}
  (\bibinfo{year}{2025}), \bibinfo{pages}{107780}.
\newblock
\urldef\tempurl%
\url{https://doi.org/10.1016/j.infsof.2025.107780}
\showURL{%
\tempurl}


\bibitem[Quiring et~al\mbox{.}(2019)]%
        {Quiring2019}
\bibfield{author}{\bibinfo{person}{Erwin Quiring}, \bibinfo{person}{Alwin
  Maier}, {and} \bibinfo{person}{Konrad Rieck}.}
  \bibinfo{year}{2019}\natexlab{}.
\newblock \showarticletitle{Misleading Authorship Attribution of Source Code
  using Adversarial Learning}. In \bibinfo{booktitle}{\emph{28th {USENIX}
  Security Symposium, {USENIX} Security 2019, Santa Clara, CA, USA, August
  14-16, 2019}}. \bibinfo{publisher}{{USENIX} Association},
  \bibinfo{pages}{479--496}.
\newblock
\urldef\tempurl%
\url{https://www.usenix.org/conference/usenixsecurity19/presentation/quiring}
\showURL{%
\tempurl}


\bibitem[Segura et~al\mbox{.}(2016)]%
        {DBLP:journals/tse/SeguraFSC16}
\bibfield{author}{\bibinfo{person}{Sergio Segura}, \bibinfo{person}{Gordon
  Fraser}, \bibinfo{person}{Ana~Bel{\'{e}}n S{\'{a}}nchez}, {and}
  \bibinfo{person}{Antonio~Ruiz Cort{\'{e}}s}.}
  \bibinfo{year}{2016}\natexlab{}.
\newblock \showarticletitle{A Survey on Metamorphic Testing}.
\newblock \bibinfo{journal}{\emph{{IEEE} Trans. Software Eng.}}
  \bibinfo{volume}{42}, \bibinfo{number}{9} (\bibinfo{year}{2016}),
  \bibinfo{pages}{805--824}.
\newblock
\urldef\tempurl%
\url{https://doi.org/10.1109/TSE.2016.2532875}
\showURL{%
\tempurl}


\bibitem[Shao et~al\mbox{.}(2024)]%
        {Shao2024}
\bibfield{author}{\bibinfo{person}{Minghao Shao}, \bibinfo{person}{Sofija
  Jancheska}, \bibinfo{person}{Meet Udeshi}, \bibinfo{person}{Brendan
  Dolan{-}Gavitt}, \bibinfo{person}{Haoran Xi}, \bibinfo{person}{Kimberly
  Milner}, \bibinfo{person}{Boyuan Chen}, \bibinfo{person}{Max Yin},
  \bibinfo{person}{Siddharth Garg}, \bibinfo{person}{Prashanth Krishnamurthy},
  \bibinfo{person}{Farshad Khorrami}, \bibinfo{person}{Ramesh Karri}, {and}
  \bibinfo{person}{Muhammad Shafique}.} \bibinfo{year}{2024}\natexlab{}.
\newblock \showarticletitle{{NYU} {CTF} Bench: {A} Scalable Open-Source
  Benchmark Dataset for Evaluating LLMs in Offensive Security}. In
  \bibinfo{booktitle}{\emph{Advances in Neural Information Processing Systems
  38: Annual Conference on Neural Information Processing Systems 2024, NeurIPS
  2024, Vancouver, BC, Canada, December 10 - 15, 2024}}.
\newblock
\urldef\tempurl%
\url{http://papers.nips.cc/paper\_files/paper/2024/hash/69d97a6493fbf016fff0a751f253ad18-Abstract-Datasets\_and\_Benchmarks\_Track.html}
\showURL{%
\tempurl}


\bibitem[Sv{\'{a}}bensk{\'{y}} et~al\mbox{.}(2021)]%
        {Valdemar2021}
\bibfield{author}{\bibinfo{person}{Valdemar Sv{\'{a}}bensk{\'{y}}},
  \bibinfo{person}{Pavel Celeda}, \bibinfo{person}{Jan Vykopal}, {and}
  \bibinfo{person}{Silvia Bris{\'{a}}kov{\'{a}}}.}
  \bibinfo{year}{2021}\natexlab{}.
\newblock \showarticletitle{Cybersecurity knowledge and skills taught in
  capture the flag challenges}.
\newblock \bibinfo{journal}{\emph{Comput. Secur.}}  \bibinfo{volume}{102}
  (\bibinfo{year}{2021}), \bibinfo{pages}{102154}.
\newblock
\urldef\tempurl%
\url{https://doi.org/10.1016/j.cose.2020.102154}
\showURL{%
\tempurl}


\bibitem[Tihanyi et~al\mbox{.}(2024)]%
        {Tihanyi2024}
\bibfield{author}{\bibinfo{person}{Norbert Tihanyi},
  \bibinfo{person}{Mohamed~Amine Ferrag}, \bibinfo{person}{Ridhi Jain},
  \bibinfo{person}{Tam{\'{a}}s Bisztray}, {and} \bibinfo{person}{M{\'{e}}rouane
  Debbah}.} \bibinfo{year}{2024}\natexlab{}.
\newblock \showarticletitle{{CyberMetric}: {A} Benchmark Dataset based on
  Retrieval-Augmented Generation for Evaluating {LLMs} in Cybersecurity
  Knowledge}. In \bibinfo{booktitle}{\emph{{IEEE} International Conference on
  Cyber Security and Resilience, {CSR} 2024, London, UK, September 2-4, 2024}}.
  \bibinfo{publisher}{{IEEE}}, \bibinfo{pages}{296--302}.
\newblock
\urldef\tempurl%
\url{https://doi.org/10.1109/CSR61664.2024.10679494}
\showURL{%
\tempurl}


\bibitem[Turtayev et~al\mbox{.}(2024)]%
        {Turtayev2024}
\bibfield{author}{\bibinfo{person}{Rustem Turtayev}, \bibinfo{person}{Artem
  Petrov}, \bibinfo{person}{Dmitrii Volkov}, {and} \bibinfo{person}{Denis
  Volk}.} \bibinfo{year}{2024}\natexlab{}.
\newblock \showarticletitle{Hacking CTFs with Plain Agents}.
\newblock \bibinfo{journal}{\emph{CoRR}}  \bibinfo{volume}{abs/2412.02776}
  (\bibinfo{year}{2024}).
\newblock
\urldef\tempurl%
\url{https://doi.org/10.48550/ARXIV.2412.02776}
\showDOI{\tempurl}
\showeprint[arXiv]{2412.02776}


\bibitem[Twist et~al\mbox{.}(2025)]%
        {llms-love-python}
\bibfield{author}{\bibinfo{person}{Lukas Twist}, \bibinfo{person}{Jie~M.
  Zhang}, \bibinfo{person}{Mark Harman}, \bibinfo{person}{Don Syme},
  \bibinfo{person}{Joost Noppen}, {and} \bibinfo{person}{Detlef~D. Nauck}.}
  \bibinfo{year}{2025}\natexlab{}.
\newblock \showarticletitle{{LLMs} Love {Python}: {A} Study of {LLMs}' Bias for
  Programming Languages and Libraries}.
\newblock \bibinfo{journal}{\emph{CoRR}}  \bibinfo{volume}{abs/2503.17181}
  (\bibinfo{year}{2025}).
\newblock
\showeprint[arXiv]{2503.17181}
\urldef\tempurl%
\url{https://doi.org/10.48550/arXiv.2503.17181}
\showURL{%
\tempurl}


\bibitem[Wan et~al\mbox{.}(2024)]%
        {Wan2024}
\bibfield{author}{\bibinfo{person}{Shengye Wan}, \bibinfo{person}{Cyrus
  Nikolaidis}, \bibinfo{person}{Daniel Song}, \bibinfo{person}{David Molnar},
  \bibinfo{person}{James Crnkovich}, \bibinfo{person}{Jayson Grace},
  \bibinfo{person}{Manish Bhatt}, \bibinfo{person}{Sahana Chennabasappa},
  \bibinfo{person}{Spencer Whitman}, \bibinfo{person}{Stephanie Ding},
  \bibinfo{person}{Vlad Ionescu}, \bibinfo{person}{Yue Li}, {and}
  \bibinfo{person}{Joshua Saxe}.} \bibinfo{year}{2024}\natexlab{}.
\newblock \showarticletitle{{CYBERSECEVAL} 3: {Advancing} the Evaluation of
  Cybersecurity Risks and Capabilities in Large Language Models}.
\newblock \bibinfo{journal}{\emph{CoRR}}  \bibinfo{volume}{abs/2408.01605}
  (\bibinfo{year}{2024}).
\newblock
\showeprint[arXiv]{2408.01605}
\urldef\tempurl%
\url{https://doi.org/10.48550/arXiv.2408.01605}
\showURL{%
\tempurl}


\bibitem[{xixiameng}(2025)]%
        {deepseek-bug}
\bibfield{author}{\bibinfo{person}{{xixiameng}}.}
  \bibinfo{year}{2025}\natexlab{}.
\newblock \bibinfo{title}{{{[Bug]DeepSeek} V3.2 fails to call tools when
  interleaved thinking is enabled}}.
\newblock
\newblock
\urldef\tempurl%
\url{https://github.com/lobehub/lobe-chat/issues/10534}
\showURL{%
\tempurl}
\newblock
\shownote{Accessed: 2026-01-27}.


\bibitem[Yang et~al\mbox{.}(2025)]%
        {Yang2025}
\bibfield{author}{\bibinfo{person}{Borui Yang}, \bibinfo{person}{Md~Afif~Al
  Mamun}, \bibinfo{person}{Jie~M. Zhang}, {and} \bibinfo{person}{Gias Uddin}.}
  \bibinfo{year}{2025}\natexlab{}.
\newblock \showarticletitle{Hallucination Detection in Large Language Models
  with Metamorphic Relations}.
\newblock \bibinfo{journal}{\emph{Proc. {ACM} Softw. Eng.}}
  \bibinfo{volume}{2}, \bibinfo{number}{{FSE}} (\bibinfo{year}{2025}),
  \bibinfo{pages}{425--445}.
\newblock
\urldef\tempurl%
\url{https://doi.org/10.1145/3715735}
\showURL{%
\tempurl}


\bibitem[Yang et~al\mbox{.}(2023a)]%
        {JohnYang2023}
\bibfield{author}{\bibinfo{person}{John Yang}, \bibinfo{person}{Akshara
  Prabhakar}, \bibinfo{person}{Karthik Narasimhan}, {and}
  \bibinfo{person}{Shunyu Yao}.} \bibinfo{year}{2023}\natexlab{a}.
\newblock \showarticletitle{{InterCode}: {Standardizing} and Benchmarking
  Interactive Coding with Execution Feedback}. In
  \bibinfo{booktitle}{\emph{Advances in Neural Information Processing Systems
  36: Annual Conference on Neural Information Processing Systems 2023, NeurIPS
  2023, New Orleans, LA, USA, December 10 - 16, 2023}}.
\newblock
\urldef\tempurl%
\url{http://papers.nips.cc/paper\_files/paper/2023/hash/4b175d846fb008d540d233c188379ff9-Abstract-Datasets\_and\_Benchmarks.html}
\showURL{%
\tempurl}


\bibitem[Yang et~al\mbox{.}(2023b)]%
        {Yang2023}
\bibfield{author}{\bibinfo{person}{John Yang}, \bibinfo{person}{Akshara
  Prabhakar}, \bibinfo{person}{Shunyu Yao}, \bibinfo{person}{Kexin Pei}, {and}
  \bibinfo{person}{Karthik~R Narasimhan}.} \bibinfo{year}{2023}\natexlab{b}.
\newblock \showarticletitle{Language Agents as Hackers: {Evaluating}
  Cybersecurity Skills with Capture the Flag}. In
  \bibinfo{booktitle}{\emph{Multi-Agent Security Workshop @ NeurIPS'23}}.
\newblock
\urldef\tempurl%
\url{https://openreview.net/forum?id=KOZwk7BFc3}
\showURL{%
\tempurl}


\bibitem[Yao et~al\mbox{.}(2023)]%
        {Yao2023}
\bibfield{author}{\bibinfo{person}{Shunyu Yao}, \bibinfo{person}{Jeffrey Zhao},
  \bibinfo{person}{Dian Yu}, \bibinfo{person}{Nan Du}, \bibinfo{person}{Izhak
  Shafran}, \bibinfo{person}{Karthik~R. Narasimhan}, {and}
  \bibinfo{person}{Yuan Cao}.} \bibinfo{year}{2023}\natexlab{}.
\newblock \showarticletitle{ReAct: Synergizing Reasoning and Acting in Language
  Models}. In \bibinfo{booktitle}{\emph{The Eleventh International Conference
  on Learning Representations, {ICLR} 2023, Kigali, Rwanda, May 1-5, 2023}}.
  \bibinfo{publisher}{OpenReview.net}.
\newblock
\urldef\tempurl%
\url{https://openreview.net/forum?id=WE\_vluYUL-X}
\showURL{%
\tempurl}


\bibitem[Zhang et~al\mbox{.}(2025b)]%
        {AndyZhang2025}
\bibfield{author}{\bibinfo{person}{Andy~K. Zhang}, \bibinfo{person}{Neil
  Perry}, \bibinfo{person}{Riya Dulepet}, \bibinfo{person}{Joey Ji},
  \bibinfo{person}{Celeste Menders}, \bibinfo{person}{Justin~W. Lin},
  \bibinfo{person}{Eliot Jones}, \bibinfo{person}{Gashon Hussein},
  \bibinfo{person}{Samantha Liu}, \bibinfo{person}{Donovan~Julian Jasper},
  \bibinfo{person}{Pura Peetathawatchai}, \bibinfo{person}{Ari Glenn},
  \bibinfo{person}{Vikram Sivashankar}, \bibinfo{person}{Daniel Zamoshchin},
  \bibinfo{person}{Leo Glikbarg}, \bibinfo{person}{Derek Askaryar},
  \bibinfo{person}{Haoxiang Yang}, \bibinfo{person}{Aolin Zhang},
  \bibinfo{person}{Rishi Alluri}, \bibinfo{person}{Nathan Tran}, {and}
  \bibinfo{person}{et al.}} \bibinfo{year}{2025}\natexlab{b}.
\newblock \showarticletitle{Cybench: {A} Framework for Evaluating Cybersecurity
  Capabilities and Risks of Language Models}. In \bibinfo{booktitle}{\emph{The
  Thirteenth International Conference on Learning Representations, {ICLR} 2025,
  Singapore, April 24-28, 2025}}. \bibinfo{publisher}{OpenReview.net}.
\newblock
\urldef\tempurl%
\url{https://openreview.net/forum?id=tc90LV0yRL}
\showURL{%
\tempurl}


\bibitem[Zhang et~al\mbox{.}(2024)]%
        {Zhang2024}
\bibfield{author}{\bibinfo{person}{Hugh Zhang}, \bibinfo{person}{Jeff Da},
  \bibinfo{person}{Dean Lee}, \bibinfo{person}{Vaughn Robinson},
  \bibinfo{person}{Catherine Wu}, \bibinfo{person}{William Song},
  \bibinfo{person}{Tiffany Zhao}, \bibinfo{person}{Pranav Raja},
  \bibinfo{person}{Charlotte Zhuang}, \bibinfo{person}{Dylan Slack},
  \bibinfo{person}{Qin Lyu}, \bibinfo{person}{Sean Hendryx},
  \bibinfo{person}{Russell Kaplan}, \bibinfo{person}{Michele Lunati}, {and}
  \bibinfo{person}{Summer Yue}.} \bibinfo{year}{2024}\natexlab{}.
\newblock \showarticletitle{A Careful Examination of Large Language Model
  Performance on Grade School Arithmetic}. In
  \bibinfo{booktitle}{\emph{Advances in Neural Information Processing Systems
  38: Annual Conference on Neural Information Processing Systems 2024, NeurIPS
  2024, Vancouver, BC, Canada, December 10 - 15, 2024}}.
\newblock
\urldef\tempurl%
\url{http://papers.nips.cc/paper\_files/paper/2024/hash/53384f2090c6a5cac952c598fd67992f-Abstract-Datasets\_and\_Benchmarks\_Track.html}
\showURL{%
\tempurl}


\bibitem[Zhang et~al\mbox{.}(2025a)]%
        {Zhang2025}
\bibfield{author}{\bibinfo{person}{Jie Zhang}, \bibinfo{person}{Haoyu Bu},
  \bibinfo{person}{Hui Wen}, \bibinfo{person}{Yongji Liu},
  \bibinfo{person}{Haiqiang Fei}, \bibinfo{person}{Rongrong Xi},
  \bibinfo{person}{Lun Li}, \bibinfo{person}{Yun Yang},
  \bibinfo{person}{Hongsong Zhu}, {and} \bibinfo{person}{Dan Meng}.}
  \bibinfo{year}{2025}\natexlab{a}.
\newblock \showarticletitle{When {LLMs} meet cybersecurity: {A} systematic
  literature review}.
\newblock \bibinfo{journal}{\emph{Cybersecur.}} \bibinfo{volume}{8},
  \bibinfo{number}{1} (\bibinfo{year}{2025}), \bibinfo{pages}{55}.
\newblock
\urldef\tempurl%
\url{https://doi.org/10.1186/s42400-025-00361-w}
\showURL{%
\tempurl}


\bibitem[Zou et~al\mbox{.}(2024)]%
        {Zou2024}
\bibfield{author}{\bibinfo{person}{Yuwen Zou}, \bibinfo{person}{Yang Hong},
  \bibinfo{person}{Jingyi Xu}, \bibinfo{person}{Lekun Liu}, {and}
  \bibinfo{person}{Wenjun Fan}.} \bibinfo{year}{2024}\natexlab{}.
\newblock \showarticletitle{Leveraging Large Language Models for Challenge
  Solving in Capture-the-Flag}. In \bibinfo{booktitle}{\emph{23rd {IEEE}
  International Conference on Trust, Security and Privacy in Computing and
  Communications, TrustCom 2024, Sanya, China, December 17-21, 2024}}.
  \bibinfo{publisher}{{IEEE}}, \bibinfo{pages}{1541--1550}.
\newblock
\urldef\tempurl%
\url{https://doi.org/10.1109/TrustCom63139.2024.00213}
\showURL{%
\tempurl}


\end{thebibliography}
